\documentclass[aps,prd,twocolumn,superscriptaddress]{revtex4-2}
\usepackage{amsmath,amssymb}
\usepackage{graphicx}
\usepackage{mathrsfs}
\usepackage{bm}
\usepackage{xcolor}
\usepackage{tikz}
\usepackage{soul} 
\usetikzlibrary{arrows.meta, positioning, shapes.geometric}
\definecolor{redlink}{RGB}{230,60,60}
\definecolor{yellowequation}{RGB}{210, 160, 60}

\usepackage[colorlinks=true,
            linkcolor=red,
            citecolor=blue,
            urlcolor=blue]{hyperref}

\begin{document}

\title{Neural posterior estimation of Galactic Binary signals for the LISA mission}

\author{Tanguy Delmond}
\affiliation{Centre National d’Etudes Spatiales – Centre spatial de Toulouse, France}
\affiliation{Univ Toulouse, ISAE-SUPAERO, ISAE-RECHERCHE, Toulouse, France}

\author{Natalia Korsakova}
\affiliation{Université Côte d'Azur, Observatoire de la Côte d'Azur, CNRS
Artemis, Bd de l'Observatoire, F-06304 Nice, France}

\author{Thomas Oberlin}
\affiliation{Univ Toulouse, ISAE-SUPAERO, ISAE-RECHERCHE, Toulouse, France}
\author{Sylvain Marsat}
\affiliation{Univ Toulouse, CNRS, L2IT, Toulouse, France}
\author{Antoine Basset}
\affiliation{Centre National d’Etudes Spatiales – Centre spatial de Toulouse, France}
\author{Nicolas Dobigeon}
\affiliation{Univ Toulouse, Toulouse INP, UT1, UTC, CNRS, IRIT, Toulouse, France}
\date{\today}


\begin{abstract}

ESA's LISA mission will open a new window onto the gravitational-wave sky by detecting signals from a wide variety of sources in the millihertz frequency band. Among these, galactic binaries are expected to be the most numerous sources observable by LISA. Their analysis and parameter estimation represent a significant challenge, as the signals are expected to strongly overlap in both the time and frequency domains.  Conventional Bayesian inference approaches, such as Markov Chain Monte Carlo sampling, are difficult to scale to this setting due to the high dimensionality of the problem and the complicated likelihood landscape which can hinder convergence. In this work, we explore simulation-based inference as a means to perform efficient parameter estimation for single galactic binaries, with a potential extension to the analysis of multiple overlapping sources. Our approach relies on a conditional normalizing flow acting as  a neural posterior estimator. The model is trained using samples generated according to a dedicated simulation framework that does not require any likelihood computation. Once trained, the neural posterior estimator enables the generation of thousands of posterior samples per second, again without explicit likelihood evaluation. We first present results  for a single source in a narrow frequency band, and then extend the analysis to wider frequency ranges. As a proof of concept, we further investigate the more challenging case of two overlapping sources. These results demonstrate the potential of likelihood-free inference as a scalable alternative to conventional Markov chain Monte Carlo sampling for the analysis of LISA galactic binaries.
\end{abstract}
\maketitle

\section{Introduction}
Since the first detection in September 2015 \cite{Abbott2016a}, gravitational waves (GW) from the coalescence of compact-object binaries are now routinely observed by ground-based interferometers such as Advanced LIGO \cite{Abbott2015a}, Advanced Virgo \cite{Acernese2014a}, and in the future by KAGRA \cite{Aso2013a}. However, these detectors are fundamentally limited at high frequencies by terrestrial noise and armlength of the detectors, which prevents the observation of GW sources with total masses much larger than a few hundred solar masses.

The Laser Interferometer Space Antenna (LISA) \cite{Colpi2024a, LISA}, a space-based GW observatory led by ESA in collaboration with NASA and scheduled for launch in the 2030s, will overcome this limitation by operating in the millihertz frequency band, well below the sensitivity range of ground-based detectors. This will enable the observation and precise characterization of a wide variety of GW sources, including massive black hole binaries (MBHBs) with masses up to millions of solar masses  \cite{Colpi2024a, Klein2016a}, stellar-mass black hole binaries, extreme mass ratio inspirals (EMRIs) \cite{Colpi2024a, Babak2017a}, stochastic GW signal from the energetic processes in the early Universe \cite{Colpi2024a, Caprini2018a} and an exceptionally large population of Galactic Binaries (GBs) mainly composed  of double white dwarf binaries but also some neutron stars or black hole binaries  \cite{GBs}. Most of these binaries will form a stochastic GW foreground \cite{Ruiter2010a, Colpi2024a}, however, up to tens of thousands will be resolved and characterized individually \cite{Kristen2023a}. This unprecedented census will enable a detailed and essentially unbiased mapping of compact binary populations across the Milky Way, providing valuable information on binary evolution, star formation history, and Galactic structure \cite{Colpi2024a, Nelemans2001a}. As a result, GBs will constitute a cornerstone of LISA’s astrophysical program and a key probe of the Milky Way in GWs.

The huge number and diversity of signals potentially observable by LISA places the data analysis at the very heart of the mission. In broad terms, gravitational-wave data analysis consists in transforming raw interferometric measurements into astrophysical information about the observed sources. This process unfolds across several key steps: the calibration and characterization of detector data, the detection and identification of gravitational-wave signals, and finally the estimation of the physical parameters of the sources like its sky localization or component masses of the system. LISA data analysis is particularly challenging due to the large and unknown number of sources overlapping in both the frequency and time domains, and the diversity of source types. In addition to the astrophysical diversity of its GW sources, LISA data will have non-stationary noise \cite{Hartwig2022a, Alvey2025a} and be affected by a complex environment effects including gaps \cite{Dey2021a}, and glitches \cite{ Muratore2025a, Castelli2025a}. Furthermore, the data stream will be dominated by GW signals, leaving virtually no signal-free intervals for estimating instrumental noise alone. The complexity of LISA data, with its high dimensionality, multiple source types, and intricate noise properties, necessitates a global fit approach \cite{Littenberg2023a, Deng2025a, Katz2025a} in which all source and noise parameters are estimated simultaneously. Implementing such an analysis requires a sophisticated pipeline, where dedicated modules for different source classes or instrumental models run in parallel and communicate throughout the process. 

Among all LISA sources, GBs present one of the most severe data analysis challenges. The core difficulty lies in the unknown number of sources and the fact that they gradually transition into the noise floor, making it extremely hard to identify sources at low SNRs — a problem commonly referred to as the confusion problem. Until now, most of the techniques proposed in the literature to estimate their parameters posterior, rely on Bayesian trans-dimensional methods, such as the Reversible-Jump Markov Chain Monte Carlo (RJMCMC) method \cite{Green1995a, Fan2010a, Littenberg2020a}. Even setting aside the trans-dimensional aspect, standard Markov Chain Monte Carlo (MCMC) approaches are computationally demanding in this context due to the cost of performing millions of likelihood evaluations, which can also lead to slow convergence. These limitations motivate the exploration of alternative Bayesian inference strategies capable of delivering accurate posterior distributions more efficiently or complementary Bayesian tools including methods that can provide fast, accurate proposals to improve the efficiency of classical MCMC and RJMCMC samplers. Bayesian inference does not necessarily have to be performed with MCMC: simulation-based inference (SBI) offers a compelling alternative, and in particular a direct route to posterior distributions over source parameters. Among SBI techniques, neural posterior estimation (NPE), through Normalizing Flows (NFs), has been widely and successfully applied to GW data analysis, demonstrating that accurate, full posterior distributions can be obtained orders of magnitude faster than with traditional samplers. In the context of ground-based detectors, the DINGO pipeline enables fast and accurate posterior estimation for compact binary mergers \cite{Green2021a, Dax2021a}. NFs have also been explored for LISA sources, including attempts at targeting MBHBs via sequential neural likelihood \cite{Vilchez2025a} and NPE \cite{AliceSparado}, as well as within the DINGO framework. These developments are partly motivated by the need for rapid parameter estimation for sources with expected electromagnetic counterparts, but the computational benefits are relevant more broadly. In parallel, work on extending SBI methods to trans-dimensional settings has begun, with approaches that start from simpler source classes and progressively address the unknown-number problem \cite{Houba2025a}. Here, we follow a complementary strategy: rather than starting from simpler toy sources, we begin by fitting a representative isolated GB signal as observed by LISA, directly targeting the full posterior distribution over the parameters of a single GB, with the goal of building toward a trans-dimensional treatment in subsequent and future works. Even at this stage, the resulting posteriors can already serve as informative proposals within classical MCMC and RJMCMC pipelines, offering a concrete near-term benefit. We therefore propose to explore the opportunity offered by NPE through NFs for accurate and efficient posterior inference on GB sources observed by LISA.

GBs are quasi-monochromatic sources, which allows them to be analyzed within very narrow frequency bands, typically of the order of a few micro hertz. Such a narrow-band approach naturally motivates the use of a Fourier-domain analysis; however, the impact of data gaps, which may challenge this representation, is beyond the scope of the present study. Hence, as a first step, our study will focus on single GB sources within narrow frequency windows. These windows will be localized on bands ranging from 1 mHz to 15 mHz as they correspond to regions of the LISA spectrum that contain the largest population of GBs. We will then explore solutions for tackling more challenging  situations, characterized by larger frequency windows or by the presence of  two overlapping sources.

This paper is organized as follows. Section~\ref{sec:sources} presents the GB sources and the LISA data. The proposed inference methodology is detailed in Section~\ref{sec:methode}, including the architecture of the neural spline to perform NPE and the generated data sets. In Section~\ref{sec:narrow_freq}, we present our results in the narrow frequency band regime and discuss the limitations of this approach. In Section~\ref{sec:Extension}, we extend our analysis to more challenging configurations. More precisely, we first extend the analysis to wider frequency bands, covering ranges larger than a few micro hertz within the LISA sensitivity band, then we focus on the case of two overlapping sources and we evaluate the performance of the model in the presence of signal superposition.

\section{LISA data model}
\label{sec:sources}

In this section, we introduce the physical framework underlying the generation of GBs and their observation by LISA. GB signals are characterized by a set of eight parameters $\theta$, which fully describe the properties of the binary. We begin by describing how these parameters contribute to the GB gravitational waveform, from the polarization amplitudes $h_+$ and $h_\times$ in the source 
frame in subsection~\ref{subsec:GWsignal}, to their expression in the Solar System Barycenter (SSB) frame in subsection~\ref{subsec:DetectorFram}. We then describe in subsection~\ref{subsec:LISAObservables} how the GW tensor 
$\boldsymbol{h}$ is projected onto the LISA constellation and processed through TDI combinations to form the orthogonal channels $A$, $E$, and $T$, with the full instrument response and TDI derivations relegated to App.~\ref{app:LISAResponse}. Finally, the observables entering our analysis pipeline are the noisy frequency-domain channels $\tilde{A}(f)$ and $\tilde{E}(f)$, defined in subsection~\ref{subsec:noise_model}.

\subsection{The GW signal}
\label{subsec:GWsignal}

Let us first introduce a source frame. In this frame, we define the gravitational-wave propagation vector $k$, pointing from the source toward the observer. We also introduce two polar angles, $(\iota, \phi_0)$. Angle $\iota$ corresponds to the inclination of the binary system, while $\phi_0$ denotes its initial phase. From this setup, we construct a basis by introducing two polarization vectors $(p, q)$. Together with the propagation direction $k$, they form an orthonormal frame $(p, q, k)$. We then construct two so-called polarization tensors
\begin{equation}
\begin{aligned}
    \boldsymbol{e}_{+} &= p \otimes p - q \otimes q \,, \\
    \boldsymbol{e}_{\times} &= p \otimes q + q \otimes p \,. \label{eq:pol_cross}
\end{aligned}
\end{equation}
In this basis, the gravitational wave tensor $\boldsymbol{h}$  in the transverse-traceless gauge with the two polarization $h_{+}$, $h_{\times}$ \cite{Maggiore} writes
\begin{equation}
    \boldsymbol{h} = h_{+}\boldsymbol{e}_{+} + h_{\times}\boldsymbol{e}_{\times} \,.
\end{equation}
GBs are in the inspiral phase and are very far from merger. We consider circular orbits, neglecting eccentricity. We then use the quadrupole formula to describe the binary waveform. The plus and cross polarizations are then given by \cite{Maggiore}
\begin{equation}
\begin{aligned}
    h_{+}(t) &=  A(1+\cos^2{(\iota}))\cos{(\Phi(t)})\,,\\
    h_{\times}(t) &= 2A\cos{(\iota})\sin{(\Phi(t)})\,,
\end{aligned}
\label{hp_hc}
\end{equation}
where $\Phi$ is the phase of the system and A the amplitude.
Since GBs are quasi-monochromatic sources, we simplify the phase equation as a Taylor expansion up to the first order in the frequency derivative
\begin{equation}
    \Phi(t) = 2\pi\left( f_{0}t+\dot{f}_{0}\frac{t^{2}}{2} \right) + \phi_{0}\,,
    \label{phase_GB}
\end{equation}
where $f_{0}$ is the frequency of the system and $\dot{f}_{0}$ the first derivative of the frequency. 

The amplitude is linked to the frequency of the signal and remains constant during the LISA observation and is given by \cite{Maggiore}
\begin{equation}
    A = 2\frac{(G\mathcal{M}_{c})^{5/3}}{c^4d_{L}}(\pi f_{0})^{2/3}\,,
\end{equation}
 where $\mathcal{M}_{c}$ is the chirp mass which is defined by the two masses of the system $m_1$ and $m_2$ via $\mathcal{M}_{c} = (m_1 m_2)^{\frac{3}{5}}/(m_1 +m_2)^{\frac{1}{5}}$ and $d_L$ is the luminosity distance. The frequency derivative is function of the chirp mass and of the frequency, assuming that the evolution of the binary is only driven by the GW emission \cite{Maggiore}
\begin{equation}
\dot{f}_0
=
\frac{96}{5}\,\pi^{8/3}
\left(\frac{G \mathcal{M}_c}{c^3}\right)^{5/3}
f_0^{11/3}\,.
\label{fdot}
\end{equation}
Knowing $A$, $f_0$, and $\dot{f}_0$, the system of equations allows us to break the degeneracy between the chirp mass and the luminosity distance, and therefore determine both $\mathcal{M}_c$ and $d_L$. However, if the frequency evolution $\dot{f}_0$ is not measured, the chirp mass and luminosity distance cannot be independently determined, as the amplitude only constrains a combination of these two parameters.

\subsection{Detector frame}
\label{subsec:DetectorFram}

We introduce a detector frame with reference vectors $(u, v)$. These vectors are orthogonal to the direction of propagation of the source $k$. The orientation of the detector frame is defined by two angles $(\lambda, \beta)$. These correspond to the sky localization of the source, given by its ecliptic latitude and longitude. We also introduce the polarization angle $\psi$. This angle represents a degree of freedom between the source frame and the detector frame. It corresponds to a rotation around the propagation vector $k$. We define the polarization tensors in the detector frame as \cite{Neil2003a}
\begin{equation}
\begin{aligned}
    \boldsymbol{\epsilon}_{+} &= u \otimes u - v \otimes v\,,  \\
    \boldsymbol{\epsilon}_{\times} &= u \otimes v + v \otimes u\,.
\end{aligned}
\end{equation}
We then can write the relation between the polarization tensors with $\psi$
\begin{equation}
\begin{aligned}
    \boldsymbol{e}_{+} &= \cos{(2\psi)}\boldsymbol{\epsilon}_{+} + \sin{(2\psi)}\boldsymbol{\epsilon}_{\times}\,,\\
    \boldsymbol{e}_{\times} &= -\sin{(2\psi)}\boldsymbol{\epsilon}_{+} + \cos{(2\psi)}\boldsymbol{\epsilon}_{\times}\,.
\end{aligned}
\end{equation}
The final expression of the GW tensor $\boldsymbol{h}$ in the detector frame is then given by:
\begin{equation}
    \boldsymbol{h} = h_{+}^{SSB}\boldsymbol{\epsilon}_{+} + h_{\times}^{SSB}\boldsymbol{\epsilon}_{\times}\,,
    \label{h_tensor}
\end{equation}
using the polarization in the SSB frame as follows:
\begin{equation}
\begin{aligned}
    h_{+}^{SSB} &=  h_{+}\cos{(2\psi)}-h_{\times}\sin{(2\psi)}\,,\\
    h_{\times}^{SSB} &= h_{+}\sin{(2\psi)}+h_{\times}\cos{(2\psi)}\,.
\end{aligned}
\label{hp_hc_SSB}
\end{equation}
Finally, each GB is parametrized by the set of the eight parameters
\begin{equation}
    \theta = \left( f_{0}, \dot{f}_{0}, \lambda, \beta, A, \iota, \phi_{0}, \psi \right)\,.
    \label{parameters}
\end{equation}

Recovering the full parameter vector $\theta$ from LISA data is the central objective of this work, and the methods developed to this end are presented in the next Section \ref{sec:methode}. However, these parameters are not directly accessible to the observer. As detailed in the following subsection, the information they encode reaches the observer only indirectly: the GW signal introduced above is first projected onto the LISA arms through the instrument response and then further processed through TDI combinations. Each of these steps introduces additional modulations and couplings between parameters, making their individual extraction a non-trivial inference problem.

\subsection{LISA observables}
\label{subsec:LISAObservables}

The GW signal is not directly accessible to the observer. The GW tensor \eqref{h_tensor} is projected onto the LISA constellation on each arm of the interferometer \cite{Neil2003a, Neil2003b, LISA2}, yielding a set of single-link measurements, each encoding the relative distance shift along one arm. To suppress the dominant laser frequency noise — several orders of magnitude above the expected GW signal — these measurements are combined into Time-Delay Interferometry (TDI) variables \cite{Massimo1999a, Neil2003c, Massimo2004a, Katz2022b}. We use first-generation TDI, in which three channels $X$, $Y$, $Z$ are constructed as linear combinations of delayed single-link measurements, canceling laser noise under the assumption of a rigid, unequal-arm constellation. The full LISA response is derived in App.~\ref{app:LISAResponse} and the explicit TDI expressions are defined in App.~\ref{app:TDI}. From $X$, $Y$, $Z$, one defines the orthogonal channels
\begin{equation}
\begin{aligned}
A(t) &= \frac{1}{\sqrt{2}} (Z(t)-X(t))\,, \\
E(t) &= \frac{1}{\sqrt{6}} (X(t) -2Y(t) + Z(t))\,, \\
T(t) &= \frac{1}{\sqrt{3}} (X(t) + Y(t) + Z(t))\,.
\end{aligned}
\label{TDIAE}
\end{equation}
The channel $T$ is insensitive to GWs in the low-frequency limit \cite{Marsat2021a} and is not used in this work. 
As mentioned above and shown in equation \eqref{phase_GB}, GBs are quasi-monochromatic signals. In the Fourier domain, their representation becomes significantly more compact, with power distributed over at most a few microhertz. We therefore adopt this representation for their analysis. We introduce then the $A$ and $E$ TDI channels in the Fourier domain as
\begin{equation}
    \tilde{A}(f) = \mathcal{F}[A(t)], \qquad \tilde{E}(f) = \mathcal{F}[E(t)]\,.
    \label{eq:observables}
\end{equation}
In this work, the $\tilde{A}$ and $\tilde{E}$ channels will serve as the final observables of the GW signal. The analysis can be equivalently formulated in terms of $\tilde{X}$, $\tilde{Y}$, and $\tilde{Z}$. The choice of the $\tilde{A}$ and $\tilde{E}$ basis is motivated by its convenience for data generation and by the availability of established tools for simulating these channels in a training-data framework.

\subsection{Noise model}
\label{subsec:noise_model}
The noise properties of the LISA instrument are characterized in the frequency domain 
by their one-sided power spectral densities (PSDs) \cite{Babak2021a}. In the simplified case of first-generation TDI and fixed armlengths, the noise can be fully characterized by the PSD of the quasi-orthogonal TDI channels $A$ and $E$. In this work, we adopt the standard LISA noise model combining proof-mass acceleration noise and optical metrology system noise, as described in the literature \cite{Babak2021a}. The corresponding PSDs of the instrumental noise components are denoted by $S_{\mathrm{pm}}(f)$ and $S_{\mathrm{op}}(f)$, and the resulting noise PSD in the orthogonal TDI channels can 
be written as $S_A(f) = S_E(f) = S_n(f)$ as follows

\begin{equation}
\begin{split}
S_n(f) =\;& 8 \sin^2\!\left(2\pi f L\right)
\\
&\Bigl[
2 \Bigl(
3 + 2 \cos(2\pi f L)
+ \cos(4\pi f L)
\Bigr) S_{\mathrm{pm}}(f)
\\
&
+ \Bigl(
2 + \cos(2\pi f L)
\Bigr) S_{\mathrm{op}}(f)
\Bigr]\,.
\end{split}
\label{PSD}
\end{equation}
The $S_{\mathrm{pm}}(f)$ and $S_{\mathrm{op}}(f)$ are defined in \cite{Colpi2024a} with their values in the science requirements documents. Under the assumption of stationary coloured Gaussian noise, the Fourier-domain TDI observables can then be expressed as the sum of signal from equation \eqref{eq:observables} and noise contributions,
\begin{equation}
\tilde{A}(f) = \tilde{A}^{\mathrm{GW}}(f) + \tilde{n}(f), \quad
\tilde{E}(f) = \tilde{E}^{\mathrm{GW}}(f) + \tilde{n}(f)\,,
\label{A_E_final}
\end{equation}
where $\tilde{A}^{\mathrm{GW}}(f)$ and $\tilde{E}^{\mathrm{GW}}(f)$ denote the noise-free TDI variables. The complex noise realization $\tilde{n}(f) = \Re(\tilde{n}(f)) + i\,\Im(\tilde{n}(f))$ 
is generated independently for each frequency bin by drawing the real and imaginary parts from independent Gaussian distributions,
\begin{equation}
\Re(\tilde{n}(f)),\ \Im(\tilde{n}(f))
\sim
\mathcal{N}\left(0,\frac{S_n(f)T_{\mathrm{obs}}}{4}\right)\,.
\label{noise}
\end{equation}

This procedure defines the noise generation process used to produce the training data. On a limited frequency band, the noise can be approximated as locally white and the white dwarf confusion noise \cite{Toubiana2024a} would be effectively degenerate with a global rescaling of the signal-to-noise ratio (SNR). The resulting noise is then added directly to the noiseless waveform in the frequency domain.

\subsection{Jaranowski–Królak–Schutz parametrization}
\label{subsec:JKS}
The two TDI variables $\tilde{A}$ and $\tilde{E}$ from equation \eqref{eq:observables} can each be mathematically rewritten in a linear form following the Jaranowski--Królak--Schutz (JKS) parametrization \cite{Whelan2014a}. In this formulation, $\tilde{A}$ and $\tilde{E}$ are expressed as linear combinations of four basis functions $\tilde{A}_{\mu}$ and $\tilde{E}_{\mu}$ depending on the parameters $\left( f_{0}, \dot{f}_{0}, \lambda, \beta \right)$, weighted by four coefficients $\mathcal{A}^{\mu}$ depending on the parameters 
$\left( A, \iota, \phi_{0}, \psi \right)$, as \cite{Whelan2014a}
\begin{equation}
\begin{aligned}
\tilde{A} &= \mathcal{A}^{\mu}(A, \iota, \phi_{0}, \psi)\,\tilde{A}_{\mu}(f_{0}, \dot{f}_{0}, \lambda, \beta)\,, \\
\tilde{E} &= \mathcal{A}^{\mu}(A, \iota, \phi_{0}, \psi)\,\tilde{E}_{\mu}(f_{0}, \dot{f}_{0}, \lambda, \beta)\,.
\end{aligned}
\label{A_E_JKS}
\end{equation}

We define the expression of all these quantities in App.~\ref{app:JKS}. This factorization separates the eight parameters $\theta$ \eqref{parameters} into two decoupled groups: the so-called amplitude parameters $\left( A, \iota, \phi_{0}, \psi 
\right)$, which enter linearly through the coefficients $\mathcal{A}^{\mu}$, and the so-called phase parameters $\left( f_{0}, \dot{f}_{0}, \lambda, \beta \right)$, which govern the structure of the signal through the basis functions $\tilde{A}_{\mu}$ and $\tilde{E}_{\mu}$. This structure can be exploited in two ways. First, it enables accelerated data generation through the hybrid on-the-fly strategy described in subsection~\ref{sec:dataset}. Second, it suggests estimating the coefficients $\mathcal{A}^{\mu}$ directly rather than 
the physical parameters $\left( A, \iota, \phi_{0}, \psi \right)$; we investigate this alternative parametrization in App.~\ref{app:NPE_JKS}.

\section{Method}
\label{sec:methode}

\subsection{Bayesian parameter estimation problem}

The goal of this work is to infer the physical parameters $\theta$ defined by \eqref{parameters} associated with a GB source from the  LISA measurements. These measurements are defined as the two noisy TDI \eqref{A_E_final}, namely
\begin{equation}
    d = \left( \tilde{A}(\theta_0), \tilde{E}(\theta_0) \right)\,,
    \label{data}
\end{equation}
where $\theta_0$ refers to the true parameters to be recovered. Within a Bayesian framework, this inference problem consists in estimating the posterior distribution 
\begin{equation}
    p(\theta|d) = \frac{p(d|\theta)p(\theta)}{p(d)}\,.
    \label{posterior}
\end{equation}
In \eqref{posterior} the prior distribution $p(\theta)$ encodes any prior knowledge about the source parameters and will be explicited in subsection \ref{sec:dataset}. The likelihood function $p(d  |\theta)$  measures the agreement between the model waveform and the observed data, as described in section \ref{sec:sources}. More precisely, for each GW channel, under the assumption of the additive Gaussian noise \eqref{noise} whose PSD $S_n(f)$ is given by \eqref{PSD}, this likelihood function can be derived up to an additive constant as
\begin{equation}
\begin{aligned}
\ln p(d|\theta) = -\frac{1}{2} \Big[
&(\tilde{A}(\theta) - \tilde{A}(\theta_{0}) \mid \tilde{A}(\theta) - \tilde{A}(\theta_{0})) \\
+\, &(\tilde{E}(\theta) - \tilde{E}(\theta_{0}) \mid \tilde{E}(\theta) - \tilde{E}(\theta_{0}))
\Big]\,,
\label{likelihood}
\end{aligned}
\end{equation}
where 
\begin{equation}
(\tilde{a} | \tilde{b}) = 4\,\Re \left[ \int_{0}^{+\infty} \frac{\tilde{a}(f)\,\tilde{b}^{*}(f)}{S_n(f)}df \right]\,
\end{equation}
denotes the standard matched-filter inner product between  two frequency-domain signals $\tilde{a}$ and $\tilde{b}$ \cite{Marsat2021a}. Because the evidence  $p(d)$ is generally intractable, the posterior distribution  \eqref{posterior} is generally explored using stochastic sampling techniques such as MCMC \cite{Littenberg2020a, Hastings1970a} or nested sampling \cite{Skilling2006a}. While these approaches provide accurate estimates of the posterior, they can become computationally expensive when repeatedly evaluating the likelihood over high-dimensional parameter spaces. As an alternative, this work proposes to derive a neural posterior surrogate, as detailed in the following subsection.

\subsection{Neural posterior estimation}

Given the data $d$ in \eqref{data}, our objective is to infer the posterior distribution $p(\theta|d)$ in \eqref{posterior} over the parameters $\theta$ without resorting to MCMC sampling. To this end, we consider a conditional generative model to construct a neural density estimator $q_{\phi}(\theta|d)$ parametrized by some trainable parameters  $\phi$, which approximates the target posterior distribution $p(\theta|d)$ \eqref{posterior}. The model is trained to minimize the discrepancy between the estimated distribution 
$q_{\phi}(\theta|d)$ and the true posterior $p(\theta|d)$. This is achieved by minimizing the Kullback-Leibler (KL) divergence between the two distributions
\begin{equation}
    \mathcal{D}_{\mathrm{KL}}\!\left(p(\theta|d) \,\|\, q_{\phi}(\theta|d)\right) 
    = \int p(\theta|d) \log \frac{p(\theta|d)}{q_{\phi}(\theta|d)} \, d\theta.
    \label{KL_p_q}
\end{equation}
Once the model has been trained, drawing posterior samples $\theta^{(n)}$ from the learned distribution $q_{\phi}(\theta|d)$ associated to a new observation $d$ requires  a single forward pass through the network. This approach thus offers a significant computational advantage over traditional methods, as the cost of inference is amortized across the training phase.

To define the surrogate distribution $q_{\phi}(\theta|d)$, we adopt a Normalizing Flow (NF) as the conditional generative model. A NF  is an invertible trainable neural network that aims to learn a distribution by transforming a given latent  distribution \cite{Rezende2015a}. More precisely, an NF learns an appropriate change of variables $ T: z \mapsto \theta = T(z)$ that approximately maps the distribution  of the latent variable $z$ towards the  distribution of the target variable $\theta$.  Since the posterior  $p(\theta|d)$ depends on the observed data $d$, the transformation must itself be conditioned on $d$, and is therefore denoted $T_d$, where the index $d$ explicitly reflects this conditioning. Finally, as $T_d$ is invertible, the density $q_{\phi}(\theta|d)$ of the transformed variable $\theta = T_d(z)$ can be expressed analytically via the change-of-variables formula. Denoting by $\pi$ the density of the latent variable $z$, the estimated posterior density is given by
\begin{equation}
    q_{\phi}(\theta | d) = \pi(T^{-1}_{d}(\theta))\left| \det J_{T_{d}}^{-1} \right|
    \label{posterior_NF}
\end{equation}
where $T_d := T_d(\,\cdot\,;\phi)$ is parametrized by the trainable weights $\phi$ and $J_{T_d}$ is the Jacobian matrix of the transformation $T_d$.

\begin{figure*}
\centering
\includegraphics[width=\textwidth]{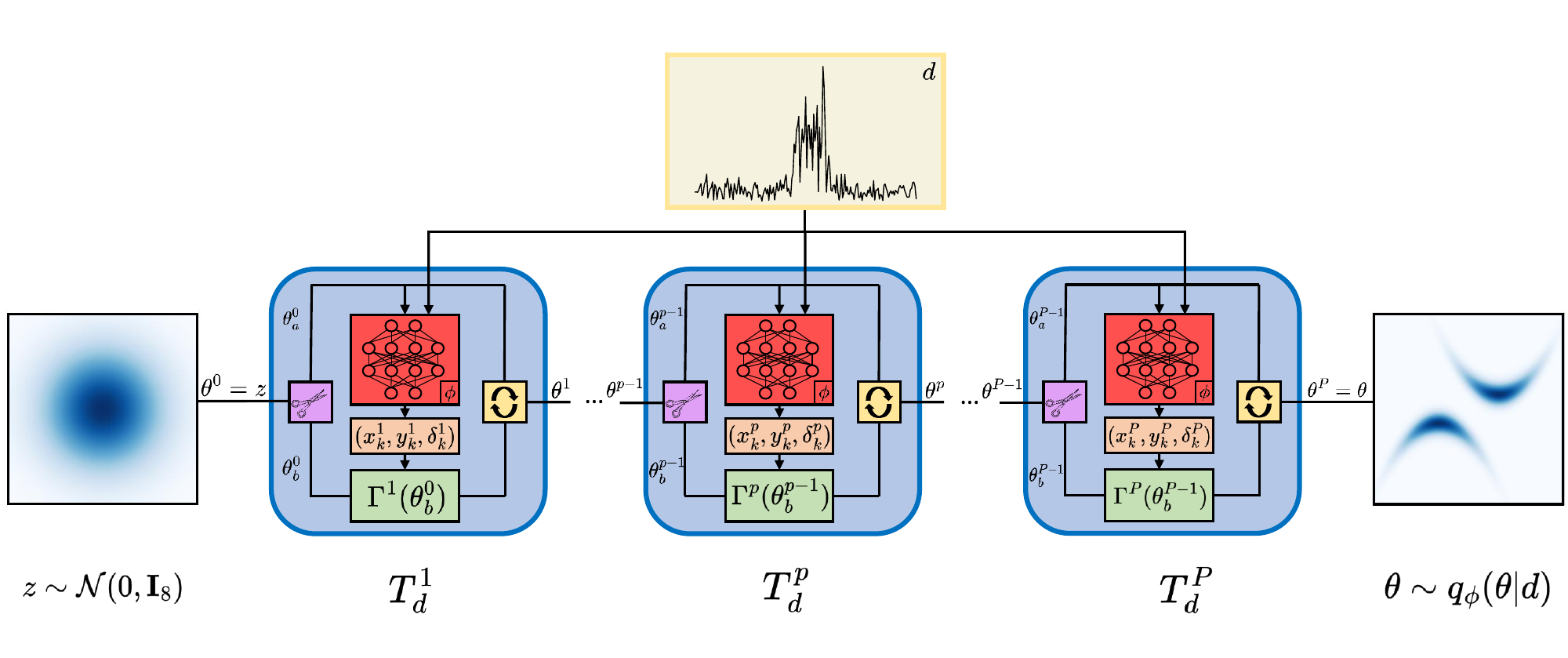}
\caption{
Architecture of the proposed data-conditioned NSF, mapping from the latent Gaussian distribution $\pi(z)$ to the neural density estimator $q_{\phi}(\theta|d)$ approximating the parameter distribution $p(\theta|d)$. Each blue block represents one flow step $T^{p}_d$ conditioned on the data $d$. The purple box represents the split of the input vector $\theta^{p-1}$ into two parts $\theta_a ^{p-1}$ and $\theta_b ^{p-1}$. The red box represents a ResNet that is parametrized by the learnable parameters $\phi$. This ResNet takes as input the variable subset $\theta_a ^{p-1}$ together with the data $d$, and predicts the spline parameters $\{(x_k^p ,y_k^p)\}_{k=0}^{K}$ and $\{\delta_k^p\}_{k=1}^{K-1}$. These parameters define the invertible transformation $\Gamma^{p}$, represented as the green box, applied to the variable subset $\theta_b ^{p-1}$. The yellow box represents the shuffling procedure applied to the variable before entering into the subsequent flow step. The arrows represent non-invertible paths, while the non-arrowed lines represent invertible paths.}
\label{NF_schem}
\end{figure*}

Regarding its architecture, an NF is composed as the sequence of $P$ invertible and differentiable transformations $T^{p}_d$ ($p=1,\ldots, P$), referred to as flow steps, namely
\begin{equation}
    T_d = T^{P}_d\circ T^{P-1}_d\circ \cdots \circ T^{1}_d\,.
\end{equation}
In particular, a widely used NF architecture is the real-valued non-volume preserving (RealNVP)  transform, which is based on affine coupling transformations \cite{Dinh2017a}. Let  $\theta^p$ denote the output of the $p$th flow
\begin{equation}
    \theta^p = T^{p}_d(\theta^{p-1})
\end{equation}
with, by definition, $\theta^0 = z$ and $\theta^P = \theta$. Under this framework, the input variables $\theta^{p-1}$ of the $p$th flow are partitioned into two subsets, i.e., $\theta^{p-1} = (\theta _a ^{p-1},\theta _b ^{p-1})$. One subset $\theta _a ^{p-1}$ remains unchanged, while the other $\theta_b ^{p-1}$ undergoes an affine transformation. To ensure a transformation of each component of the vector $\theta$, their order are shuffled after each flow step. This construction ensures that the mapping $T_d$ is bijective and differentiable, with a tractable triangular Jacobian matrix which allows efficient computation of its determinant during training. Building upon this idea, neural spline flows (NSF) replace the affine transformations of RealNVP with monotonic rational--quadratic spline transformations, providing significantly greater flexibility for modeling complex distributions \cite{Durkan2019a}. Previous studies devoted to gravitational-wave parameter estimation have demonstrated excellent performance using NSF-based architectures \cite{Dax2021a, Green2021a, AliceSparado}. We therefore adopt the NSF architecture and define the transformations as
\begin{equation}
T^{p}_d(\theta^{p-1}) =
\begin{cases}
\theta _a ^{p-1} \\
\Gamma^{p}(\theta _b ^{p-1}; \theta _a ^{p-1}, d)
\end{cases}
\end{equation}
where $\Gamma^{p}$ is defined through monotonic rational-quadratic splines on a bounded interval $[x_{\min}, x_{\max}]$. For each flow-step, the spline is parameterized by $K+1$ knots with monotically increasing coordinates $\{(x_k^p,y_k^p)\}_{k=0}^{K}$ and a set of $K-1$ derivatives $\{\delta_k^p\}_{k=1}^{K-1}$. These parameters are learned by a trainable residual neural network (ResNet), one for each flow step, whose input is the concatenation of the unchanged variables $\theta _a ^{p-1}$ and the data $d$  without any embedding step \cite{He2016a}. Each ResNet is composed of a sequence of residual blocks, and each residual block consists of a small neural network made of two linear layers with nonlinear activations, followed by a skip connection that adds the block input back to its output. By stacking multiple NSF layers as depicted in  Fig.~\ref{NF_schem}, the model is able to describe highly nonlinear and correlated structures in the posterior distribution, making it particularly well-suited for complex inference problems.

In practice, following the strategy adopted by most of the works of the literature and, as suggested by the name of the model, the base distribution $\pi(z)$ is chosen as the standard multivariate normal distribution $\mathcal{N}(0,\mathbf{I}_8)$ from which it is simple to sample from. Thus, once the neural posterior estimator has been trained, generating a sample $\theta^*$ from the trained distribution $q_{\phi}(\theta|d)\approx p(\theta|d)$ can be easily conducted by drawing a sample $z^* \sim \pi(z)$ and applying the learned transformation $\theta^* = T_d(z^*)$.

\subsection{Generation of the training set}
\label{sec:dataset}

Here we present the simulation protocol to generate the waveform dataset used for  training the proposed neural posterior model.\\

\noindent \textbf{Hybrid data generation strategy --} Training the NF requires a large set of simulated LISA observations $d$ with associated GB parameters $\theta$. Each training sample indexed by $n$ thus consists of a pair $(\theta^{(n)}, d^{(n)})$ $(n\geq 1)$, where the source parameter $\theta^{(n)}$ is drawn from the prior distribution $\theta^{(n)} \sim p(\theta)$ (discussed later), and the waveform $d^{(n)}$ is conditionally drawn from the distribution $d^{(n)} \sim p(d|\theta^{(n)})$. Although the training dataset could in principle be generated either fully on-the-fly during training or entirely precomputed and stored beforehand, we adopt an intermediate strategy that provides a trade-off between waveform diversity and computational efficiency. More specifically, we exploit the equivalent linear expansion described in subsection~\ref{subsec:JKS} to implement a hybrid on-the-fly generation procedure.  A predefined set of $N = 7 \times 10^6$ quadruplet functions $\tilde{A}_{\mu}^{(n)}$ and $\tilde{E}_{\mu}^{(n)}$ is first generated according to \eqref{A_E_JKS} by randomly drawing $N$ samples  $f_{0}^{(n)}$, $\dot{f}_{0}^{(n)}$, $\lambda^{(n)}$ and $ \beta^{(n)}$  of the first four parameters  from their respective prior distributions ($n=1,\ldots,N$). During training, the remaining parameters are generated on-the-fly according to the following procedure: \emph{i)} a quadruplet pair $(\tilde{A}_{\mu}^{(n)}, \tilde{E}_{\mu}^{(n)})$ is uniformly drawn from the set computed in advance, \emph{ii)} samples  $A^{(n)}$, $\iota^{(n)}$, $\phi_{0}^{(n)}$ and $\psi^{(n)}$ of the four remaining parameters  are drawn from their respective prior distributions, \emph{iii)} the weighting coefficients $\mathcal{A}^{\mu(n)}$ are computed according to \eqref{A_E_JKS}, \emph{iv)} the response is obtained using \eqref{A_E_JKS}, and \emph{v)} the noise-free signal is constructed from \eqref{data} by concatenating the real and imaginary parts of the $\tilde{A}$ and $\tilde{E}$ channels. The SNR of the waveform is then calculated and we retain only those with an SNR between 10 and 100 in order to focus on detectable signals while covering a broad range of physically relevant cases. Finally, instrumental noise is added on fly to the data, sampled from the LISA power spectral density \eqref{PSD} through equations \eqref{A_E_final} and \eqref{noise}. Adopting this hybrid generation strategy allows the generated dataset $\{(\theta^{(n)}, d^{(n)})\}_{n \geq 1}$ to be arbitrarily large given a finite set of $N$  quadruplet pairs $(\tilde{A}_{\mu}^{(n)}, \tilde{E}_{\mu}^{(n)})$, thus reducing the overall computational burden significantly. 

Another aspect of the equivalent linear expansion described in subsection~\ref{subsec:JKS} is that, since there is a bijection between the parameters $\left( A, \iota, \phi_{0}, \psi \right)$ and the parameters $\mathcal{A}^\mu$ \eqref{A_E_JKS}, these parameters can be estimated instead of the physical parameters. Since their posterior distribution is uni-modal and therefore expected to be simpler, we trained networks to estimate them. However, we found that the results were worse. We present this study in App.~\ref{app:NPE_JKS}.\\

\noindent \textbf{Generating GB waveforms --} For simulating GB waveforms, we use the \texttt{GBGPU} code \cite{Katz2022a}, which is itself based on \texttt{FastGB} \cite{Neil2003a} based on a C code. All signals are simulated assuming one year of observation. 
We briefly present the signal heterodyning method implemented in \texttt{GBGPU}, which enables fast computation in App.~\ref{GBGPU}. All the implementation is based on the LISAFlow code~\cite{lisaflow}, an adaptation of the the Neural Spline Flow code to LISA and GWs.

\begin{figure}
\centering
\includegraphics[width=\columnwidth]{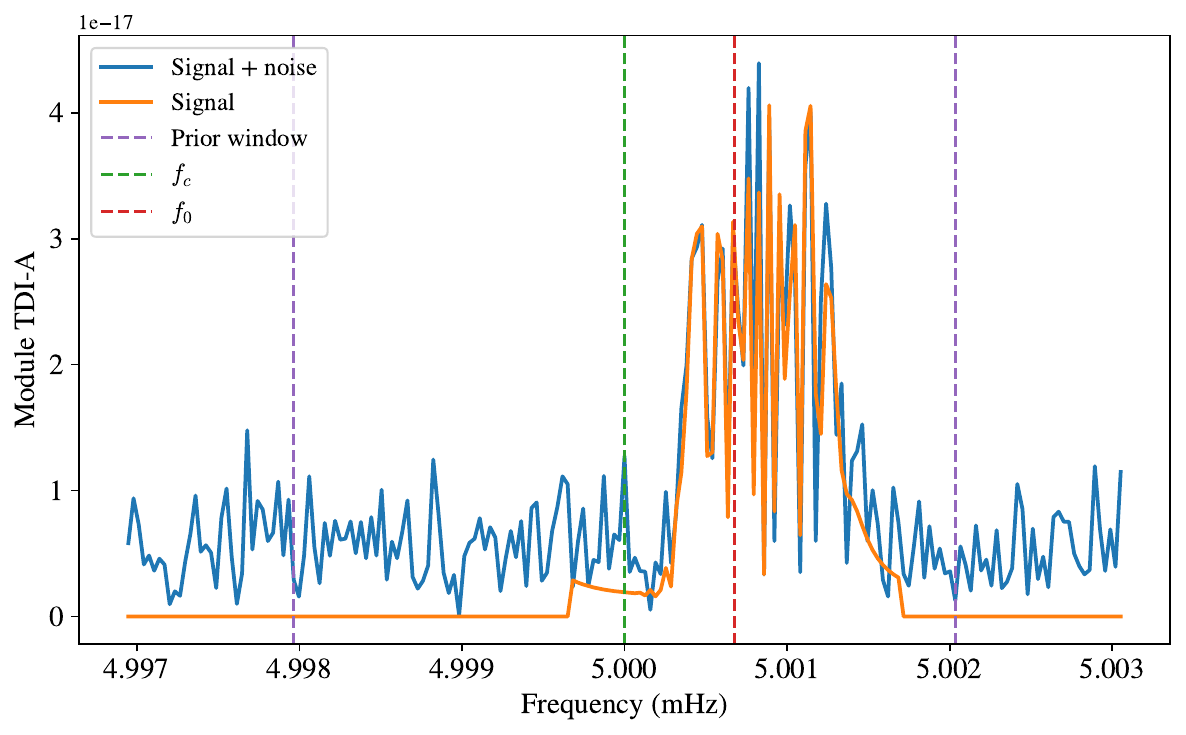}
\caption{\label{fig:gb} Examples of noise-free (orange) and noisy (blue) GB waveforms. The GB frequency $f_0$ (dashed red line) has been randomly drawn from a prior distribution defined on a  $4 \mu$Hz-width frequency window (purple dashed lines). The GB waveform is located on a $6 \mu$Hz-width frequency window. Both frequency windows are centered around the central frequency $f_{\mathrm{c}} = 5$mHz (dashed green line).}
\end{figure}

GBs are quasi-monochromatic sources; therefore, in the frequency domain, they can be analyzed in a narrow frequency window. In the frequency domain, their bandwidth arises from the Doppler effect due to LISA motion, as well as from the intrinsic frequency derivative parameter $\dot{f}_0$. The Doppler effect depends directly on the source frequency $f_0$ and on the ecliptic latitude $\beta$, so that higher-frequency sources exhibit a larger bandwidth. This bandwidth can approximately range from $0.25\,\mu\text{Hz}$ to $2\,\mu\text{Hz}$. Each waveform of the dataset is thus defined on very narrow frequency window characterized by a central frequency denoted as $f_{\mathrm{c}}$. The admissible range of the GB frequency $f_0$ is then set as $f_{\mathrm{c}}\pm 2\,\mu\text{Hz}$. Finally, the waveform is located in a wider window of size $3\,\mu\text{Hz}$ around the central frequency $f_{\mathrm{c}}$, as depicted in Fig.~\ref{fig:gb}.\\

\noindent \textbf{Parameter priors --} The GB parameters in $\theta = \left( f_{0}, \dot{f}_{0}, \lambda, \beta, A, \iota, \phi_{0}, \psi \right)$ are assumed to be a priori independent and distributed according to uniform priors defined over the ranges reported in Table~\ref{tab:priors}. As the frequency derivative $\dot{f}_0$ depends on $f_0$ according to \eqref{fdot}, its prior is adjusted according to the value of the central frequency $f_{\mathrm{c}}$. We use a uniform prior whose lower bound $\dot{f}_0^\mathrm{\text{ }min}$ and upper bound $\dot{f}_0^\mathrm{\text{ }²max}$ are taken from the Mojito catalog \cite{Toubiana2024a}. In practice, to improve the generalization ability of  the proposed neural posterior model, during the training stage, the values of the GB parameters have been normalizing  by subtracting the prior mean and  dividing by their prior standard deviation. Consequently, during the inference stage, the corresponding de-normalization should be applied to the samples generated by the trained neural posterior model.

\subsection{Training procedure}

The objective of the training is to make the neural posterior estimator $q_{\phi}(\theta|d)$ as close as possible as the true posterior $p(\theta|d)$. This is achieved by minimizing equation \eqref{KL_p_q} over the $\phi$ parameters, leading to the optimization problem
\begin{equation}
\min_{\phi}
\mathcal{D}_{\mathrm{KL}}\!\left(p(\theta|d)\,\|\,q_{\phi}(\theta|d)\right).
\label{opt_KL}
\end{equation}
Since the term $p(\theta|d)$ does not depend on the model parameters $\phi$, minimizing the KL divergence is equivalent to minimizing the negative expected log-likelihood of the estimator under the true posterior, leading to the loss function \cite{Dax2021a} \cite[Chap. 5]{Goodfellow-et-al-2016}
\begin{equation}
\mathscr{L}(\phi)
=
\mathbb{E}_{\theta \sim p(\theta)}
\mathbb{E}_{d\sim p(d|\theta)}
\left[
-\log q_{\phi}(\theta|d)
\right]\,.
\end{equation}
However this loss is untractable and we can use instead the Monte Carlo approximation on a batch to evaluate it, i.e.,  
\begin{equation}
    \mathscr{L}(\phi) \simeq -\frac{1}{M}\sum_{m=1}^{M} \log{q_{\phi}(\theta^{(m)}|d^{(m)})}
\end{equation}
where $M$ denotes the batch size and the samples $\theta^{(m)}\sim p(\theta)$ and $d^{(m)}\sim p(d|\theta)$ are drawn from the dataset generated beforehand (see subsection \ref{sec:dataset}). We can then calculate the gradient of the loss and update the network parameters $\phi$ using the Adam optimizer \cite{ADAM}.

\begin{table}
\caption{\label{tab:priors}
Ranges of the uniform parameter prior distributions used for generating the waveforms.}
\begin{ruledtabular}
\begin{tabular}{lcc}
 \textbf{Parameter} & \textbf{Lower bound} & \textbf{Upper bound} \\
\hline
$f_0 \; [\mathrm{mHz}]$        & $f_{\mathrm{c}} - 2\,\mu\mathrm{Hz}$ & $f_{\mathrm{c}} + 2\,\mu\mathrm{Hz}$ \\
$\dot{f}_0 \; [\mathrm{Hz\,s^{-1}}]$        & $\dot{f}_0^\mathrm{\text{ }min}$ & $\dot{f}_0^\mathrm{\text{ }max}$ \\
$\lambda \; [\mathrm{rad}]$    & $0$ & $2\pi$ \\
$\sin\beta$                    & $-1$ & $1$ \\
$A$                             & $10^{-24}$ & $5\times10^{-22}$ \\
$\cos\iota$                    & $-1$ & $1$ \\
$\phi_0 \; [\mathrm{rad}]$     & $0$ & $2\pi$ \\
$\psi \; [\mathrm{rad}]$       & $0$ & $\pi$ \\
\end{tabular}
\end{ruledtabular}
\end{table}

\begin{figure}
\centering
\includegraphics[width=\columnwidth]{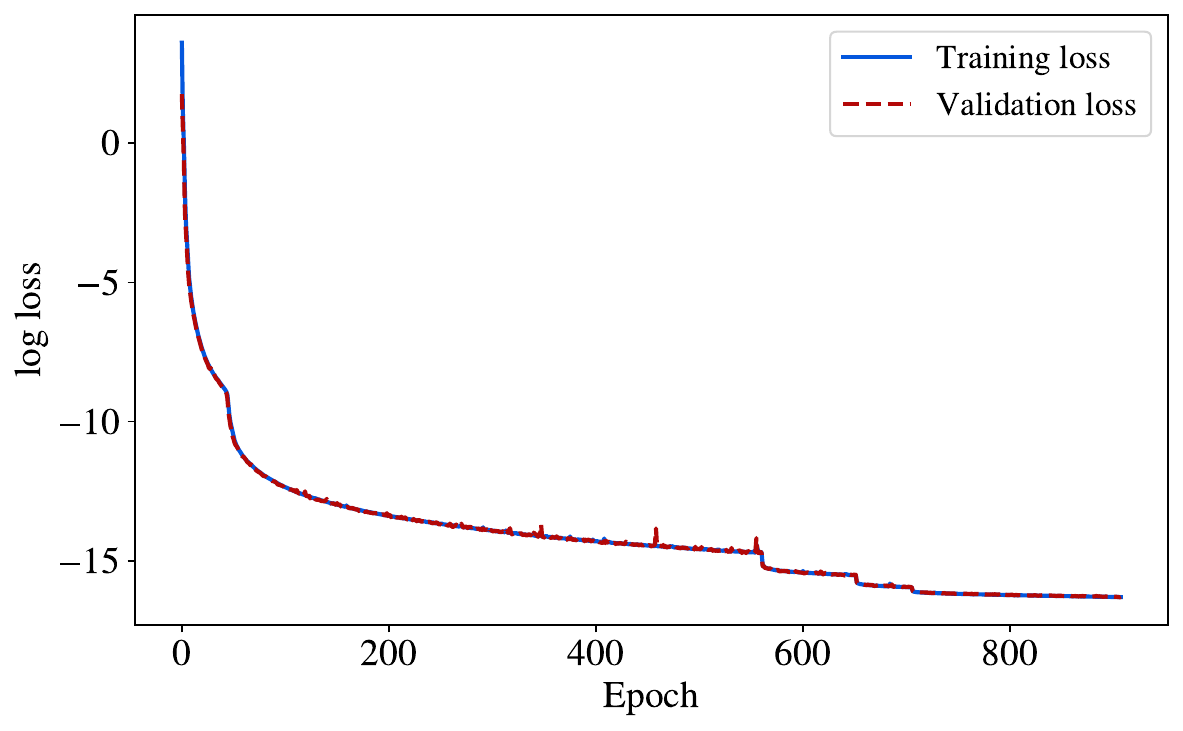}
\caption{\label{fig:loss} Training (blue) and validation (red) losses as functions of training epochs. The observed drops in the losses coincide with scheduled reductions of the learning rate.}
\end{figure}

For all trained networks, we used $1\times 10^{6}$ waveforms for the validation process. Training procedures converged after hundreds of epochs with a batch size of 2048. The learning rate was set to $2 \times 10^{-4}$ and subsequently reduced by a factor two once the validation loss had not decreased after 10 epochs thanks to the \texttt{ReduceLROnPlateau} scheduler \cite{scheduler}. The convergence required approximately one week for each training on a NVIDIA A100 GPU. We used the validation dataset to monitor the overfitting of the network on the training dataset. We never observed any overfitting. One typical example of training and validation losses is depicted in Fig.~\ref{fig:loss}.

\section{Experiments}
\label{sec:narrow_freq}

This section introduces our experimental protocol and performance metrics, and then presents and discusses our main benchmark and results.
Supplementary experiments we have carried out will be discussed in the next section.

\subsection{Experimental protocol}

We conducted a series of experiments designed to explore the parameter space of GBs. In particular, these experiments have been conducted for several values of the central frequency defining the analysis window, namely $f_{\mathrm{c}} \in \left\{1, 2, 5, 6, 10, 15\right\}$ [$\mathrm{mHz}$]. As explained in subsection~\ref{sec:dataset}, the bandwidth of the waveform increases with the frequency. Therefore, the number of spectral bins needed to correctly describe the waveform also increases with $f_{\mathrm{c}}$. To accommodate for this dependency, let divide the analysis into two cases: the low-frequency regime corresponding to central frequencies $f_{\mathrm{c}} \in \left\{1, 2, 5, 6\right\}$ [$\mathrm{mHz}$], and the high-frequency regime, corresponding to central frequencies $f_{\mathrm{c}} \in \left\{10, 15\right\}$ [$\mathrm{mHz}$]. For the low-frequency regime, only $64$ frequency bins are needed to describe the waveform for each channel. Conversely, for the high-frequency regime, more frequency bins are needed, up to $512$ points per channel.

To account for the variable length of the input waveform $d$, in this work, we do not employ any strategy of data embedding. Instead, a specific neural posterior model is trained for each considered value of the central frequency $f_{\mathrm{c}}$. Moreover the model architecture is slightly adapted with respect to the characteristics of the input waveform. This choice is motivated by our objective of isolating and quantifying potential biases arising solely from the network architecture, without introducing additional inductive biases that could result from a learned embedding module. Moreover, the frequency-domain representation of GB data is already highly compressed by construction, reducing the need for further input-level feature extraction. Thus, for all the values of the central frequency $f_{\mathrm{c}}$ considered hereafter, we use the same backbone architecture depicted in Fig.~\ref{NF_schem}. We choose $P=12$ flow steps, and for each flow step, the ResNet predicts $K=10$ bins for the spline transformation. Each ResNet is composed of $10$ residual blocks, each with ReLU activation functions. Then the only parameter that differs between the two frequency regimes is the number of neurons per layer. More precisely, in the low-frequency regime, we use $512$ neurons per layer, corresponding to a total of $1.2 \times10^8$ trainable network parameters. Conversely, since higher-frequency signals are expected to exhibit richer and more complex structures, a larger number of network parameters is adopted in the high-frequency regime. Thus, in this case, we use $1024$ neurons per layer, corresponding to a total of $3.9 \times10^8$ network parameters.

Once the neural posterior model has been trained for a specific value of the central frequency $f_{\mathrm{c}}$, it can be used  to draw sample  from the learned distribution. More precisely, for each test waveform $d$ located around $f_{\mathrm{c}}$ and analyzed hereafter, we draw $10^{4}$ samples from the trained neural posterior estimator $q_{\phi}(\theta|d)$, which requires less than one second with a computing node equipped with a NVIDIA A100 GPU. We compare the generated samples to those obtained from a classical Bayesian inference method, namely an MCMC sampler exploiting a Metropolis–Hastings algorithm based on \texttt{ptemcee} \cite{Foreman-Mackey2013a, Vousden2016a} and some parallel tempering. This MCMC algorithm targets the same distribution \eqref{posterior} as the one approximated by the proposed neural posterior model, i.e., it is defined by the likelihood \eqref{likelihood} and the prior distributions specified in Table~\ref{tab:priors}. To ensure a fair comparison of the performance reached by the two methods, all waveforms whose SNRs are not in the range $10$ -- $100$ are rejected from the proposal. Under this setup, the MCMC algorithm typically requires up to about one hour to reach convergence and to produce a same amount of posterior samples.

\begin{table}
\caption{\label{tab:waveform_params_low}
Low-frequency case: true waveform parameters for two SNR configurations.}
\begin{ruledtabular}
\begin{tabular}{lcc}
 \textbf{Parameters} & \textbf{SNR = 11} & \textbf{SNR = 60}\\
\hline
$f_0 \; [\mathrm{mHz}]$        & $0.9984$ & $4.9986$ \\
$\dot{f}_{0} \; [\mathrm{Hz\,s^{-1}}]$& $4.48 \times10^{-18}$ & $1.47 \times10^{-15}$ \\
$\beta \; [\mathrm{rad}]$     & $-3.54$ & $0.13$ \\
$\lambda \; [\mathrm{rad}]$   & $5.31$ & $5.23$ \\
$A$                           & $1.04 \times10^{-22}$ & $5.17 \times10^{-23}$ \\
$\iota \; [\mathrm{rad}]$     & $0.82$ & $1.03$ \\
$\phi_0 \; [\mathrm{rad}]$    & $0.10$ & $3.52$ \\
$\psi \; [\mathrm{rad}]$      & $1.53$ & $1.08$ \\
\end{tabular}
\end{ruledtabular}
\end{table}

\subsection{Low-frequency case}
\label{sec:low_frequency_case}

To assess the quality of the samples produced by the neural posterior estimator in the low-frequency regime, we consider two experimental configurations: a waveform around $f_{\mathrm{c}} = 1$ mHz with a low SNR of 11 and a waveform around $f_{\mathrm{c}} = 5$ mHz with a higher SNR of 60. The respective waveform parameters, drawn from the prior distributions in Table~\ref{tab:priors}, are reported in Table~\ref{tab:waveform_params_low}.  The estimation results are represented as corner plots depicting the empirical one- and two-dimensional marginal posterior distributions obtained from the samples generated by the proposed neural posterior model, alongside the corresponding empirical distributions obtained by the MCMC method.

\begin{figure*}
\centering
\includegraphics[width=\textwidth]{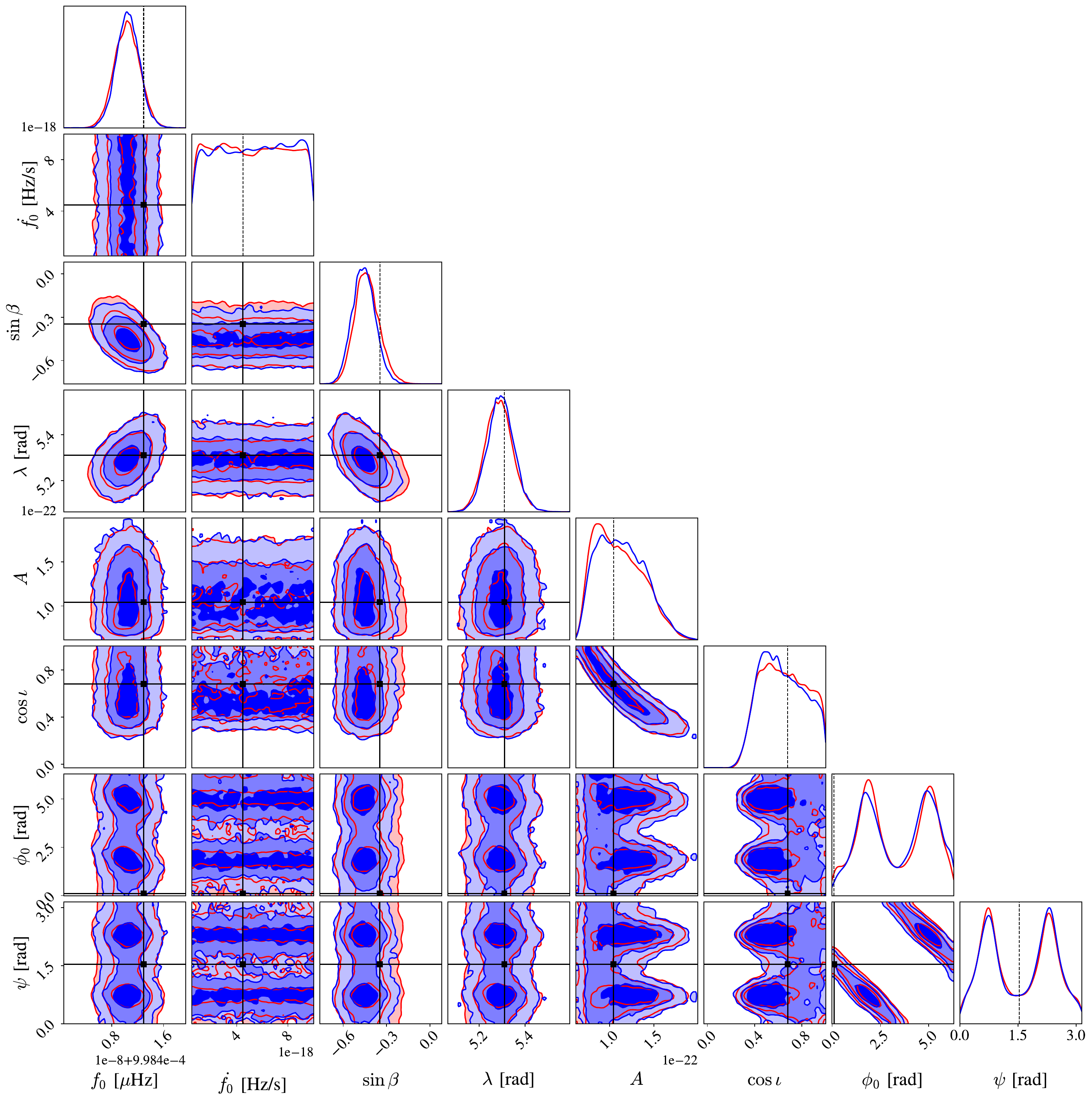}
\caption{
Low-frequency low-SNR case: empirical one- and two-dimensional marginal posterior distributions obtained by the proposed neural posterior estimator (blue) and the MCMC sampling method (red). The contours correspond to the 1$\sigma$, 2$\sigma$, and 3$\sigma$ credible regions, while the dark lines indicate the true parameter values.
}
\label{fig:SNR11}
\end{figure*}

Fig.~\ref{fig:SNR11} displays the posterior distributions for the low-SNR case. We observe very good agreement between the two approaches. The neural posterior model successfully captures the complex multimodal structure of the posterior distributions of parameters $\phi_0$ and $\psi$, as well as the non-trivial correlations between parameters ${A, \iota, \phi_0, \psi}$.

\begin{figure*}
\centering
\includegraphics[width=\textwidth]{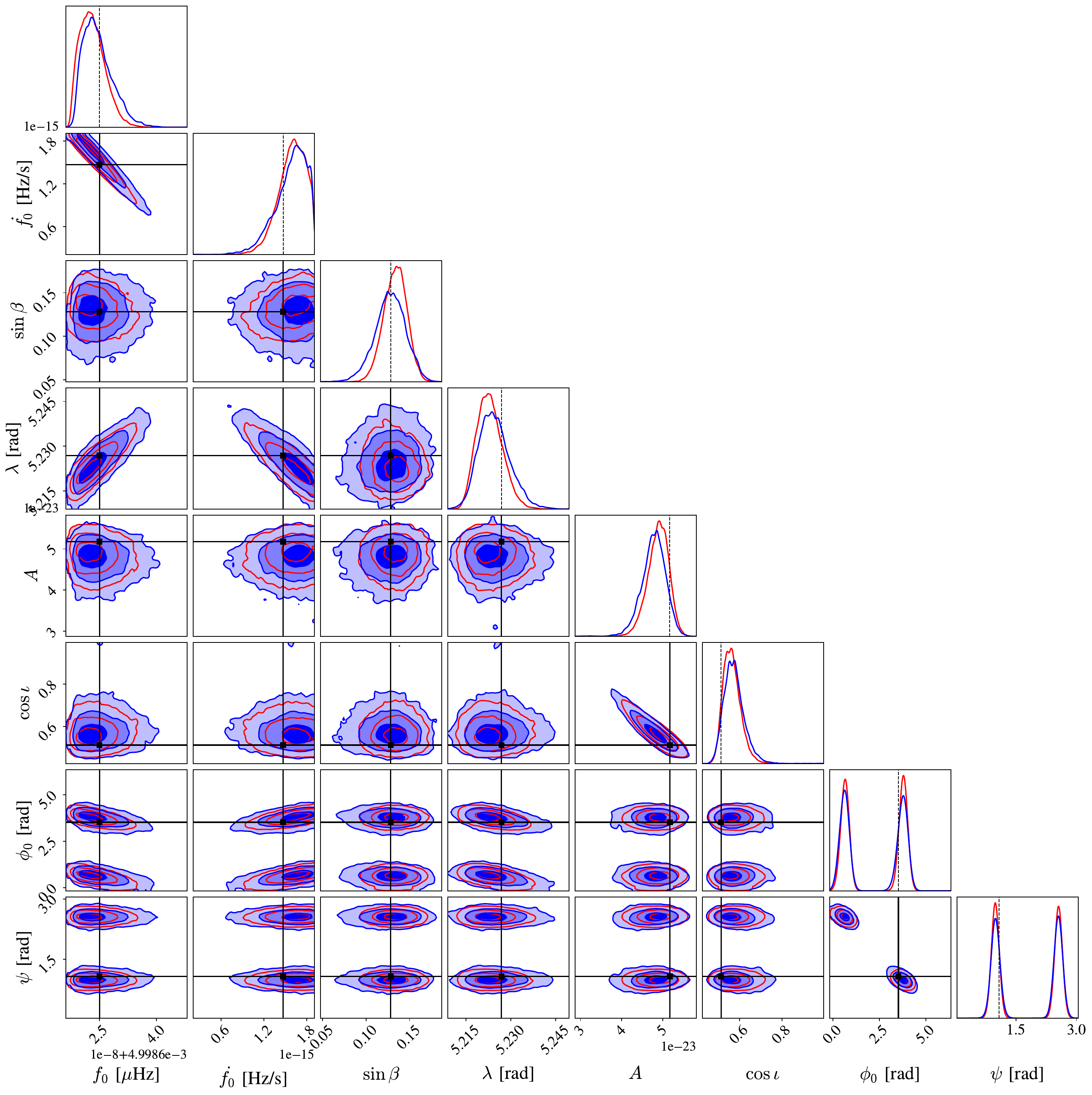}
\caption{
Low-frequency high-SNR case: empirical one- and two-dimensional marginal posterior distributions obtained by the proposed neural posterior estimator (blue) and the MCMC sampling method (red). The contours correspond to the 1$\sigma$, 2$\sigma$, and 3$\sigma$ credible regions, while the dark lines indicate the true parameter values.
}
\label{fig:SNR60}
\end{figure*}

In Fig.~\ref{fig:SNR60}, we show the posterior distributions obtained in the high-SNR case. We again observe good agreement between the two approaches. In particular the proposed neural posterior estimator successfully recovers the overall shape and structure of the posterior distributions. However, some minor differences can be noticed, particularly for the parameters $f_0$, $\dot{f}_0$, $\beta$ and $\lambda$, whose  marginal two-dimensional distributions appear slightly broader than those obtained by the MCMC sampling procedure. These small discrepancies are likely related to the higher central frequency considered in this configuration ($f_{\mathrm{c}}=5\,\mathrm{mHz}$ compared to $f_{\mathrm{c}} =1\,\mathrm{mHz}$ in the previous case). It leads to a more complex waveform structure in the frequency domain and a correspondingly more challenging inference problem, despite the overall good approximation of the posterior distribution.

\begin{figure}
\centering
\includegraphics[width=\columnwidth]{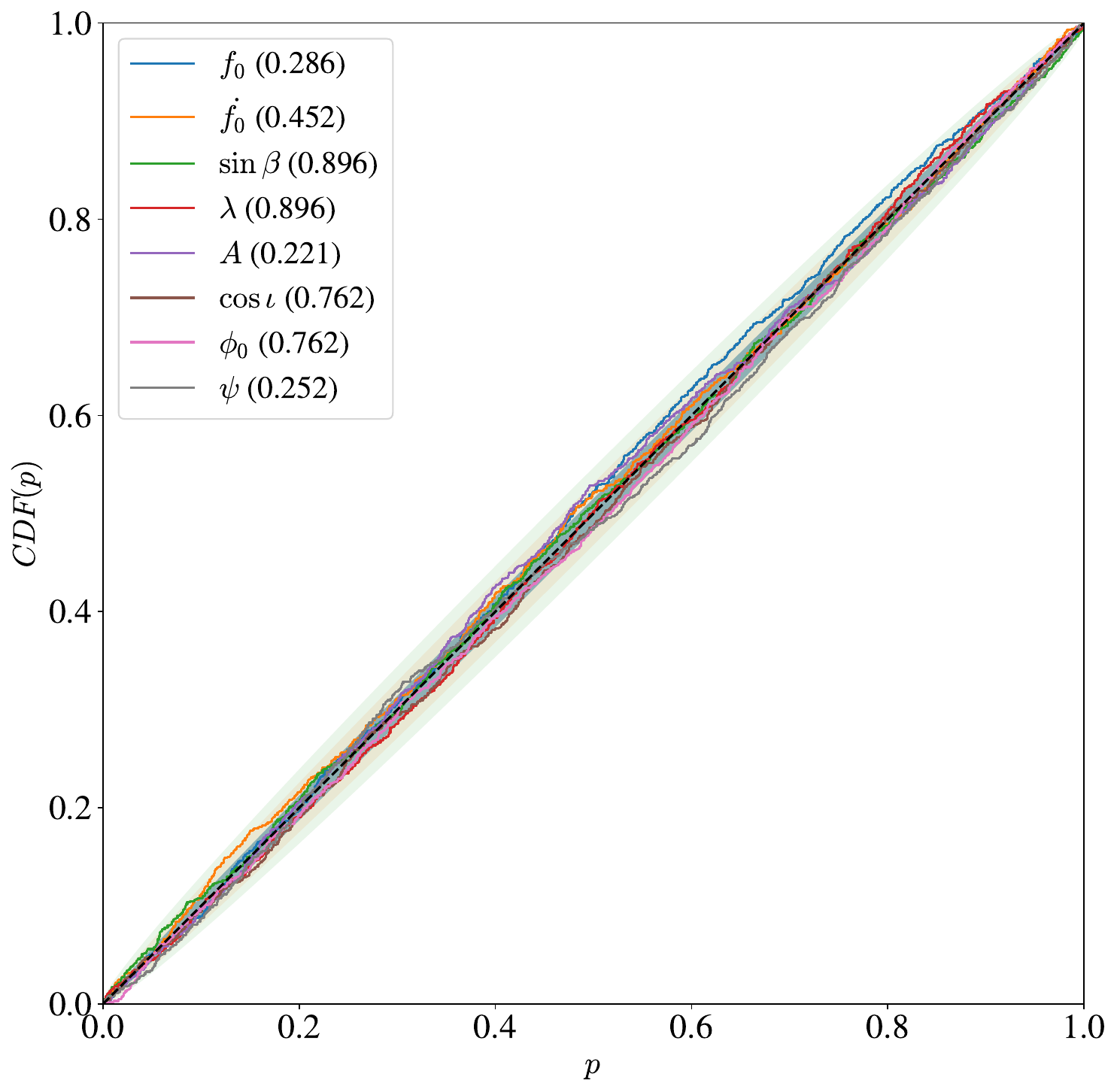}
\caption{\label{fig:ppplot} Probability-probability (p-p) plot for 1000 simulated waveforms analyzed with the flow-based network. For each injection and each one-dimensional marginalized posterior, the percentile rank of the injected parameter is computed. Colored curves show the cumulative distribution functions (CDFs) of the recovered percentiles for all parameters. The dashed diagonal corresponds to the expected uniform distribution for a perfectly calibrated model, while the shaded regions indicate the expected $1\sigma$, $2\sigma$, and $3\sigma$ confidence intervals. Kolmogorov--Smirnov (KS) test $p$-values are reported in the legend.}
\end{figure}

The statistical calibration of the neural posterior model with respect to the MCMC sampling approach is further assessed by depicting a probability-probability (p-p) plot, i.e., the cumulative distribution function (CDF) of the percentile ranks of the true parameters within the empirical one-dimensional marginal posteriors.  For a perfectly calibrated inference model, these percentiles are expected to follow a uniform distribution between $0$ and $1$, i.e., with a p-p curve defined as the diagonal between the points $(0,0)$ and $(1,1)$. Fig.~\ref{fig:ppplot} displays the recovered CDFs associated with the eight parameters, obtained from a set of 1000 waveforms drawn from the prior distributions in Table~\ref{tab:priors} in the case of the central frequency $f_{\mathrm{c}}=1$ mHz. The close agreement of the CDFs with the diagonal indicates that the trained neural posterior model accurately samples from the target posterior distributions without significant bias. This behavior is further supported by a Kolmogorov-Smirnov test, which does not reject the null hypothesis of uniformity at the 95\% confidence level.

\subsection{High-frequency case}
In the high-frequency regime, the inference task becomes more complicated because of the increased complexity of the signal morphology. As a result, the performance of the neural posterior estimator has been shown to be generally less accurate than the ones obtained in the low-frequency regime. Contrary to the analysis conducted in subsection \ref{sec:low_frequency_case}, we therefore do not  thoroughly investigate  the impact of the SNR on the performance. Instead, we  consider a unique waveform corresponding to a central frequency $f_{\mathrm{c}}=10$ mHz and with parameters drawn from the prior distributions specified  in Table~\ref{tab:priors} and reported in Table~\ref{tab:waveform_params_high}. For this single waveform, we concentrate on the overall quality of the estimated posterior distributions and, in particular, on their width and ability to recover the source parameters.

\begin{table}
\caption{\label{tab:waveform_params_high}
How-frequency case: true waveform parameters. }
\begin{ruledtabular}
\begin{tabular}{lclc}
\textbf{Parameters} & & \textbf{Parameters} & \\
\hline
$f_0 \; [\mathrm{mHz}]$                 & $10.0012$             & $A$                        & $2.98 \times10^{-23}$ \\
$\dot{f}_{0} \; [\mathrm{Hz\,s^{-1}}]$ & $7.85 \times10^{-15}$ & $\iota \; [\mathrm{rad}]$  & $2.47$ \\
$\beta \; [\mathrm{rad}]$               & $0.87$                & $\phi_0 \; [\mathrm{rad}]$ & $3.44$ \\
$\lambda \; [\mathrm{rad}]$             & $3.87$                & $\psi \; [\mathrm{rad}]$   & $0.03$ \\
\end{tabular}
\end{ruledtabular}
\end{table}

\begin{figure*}
\centering
\includegraphics[width=\textwidth]{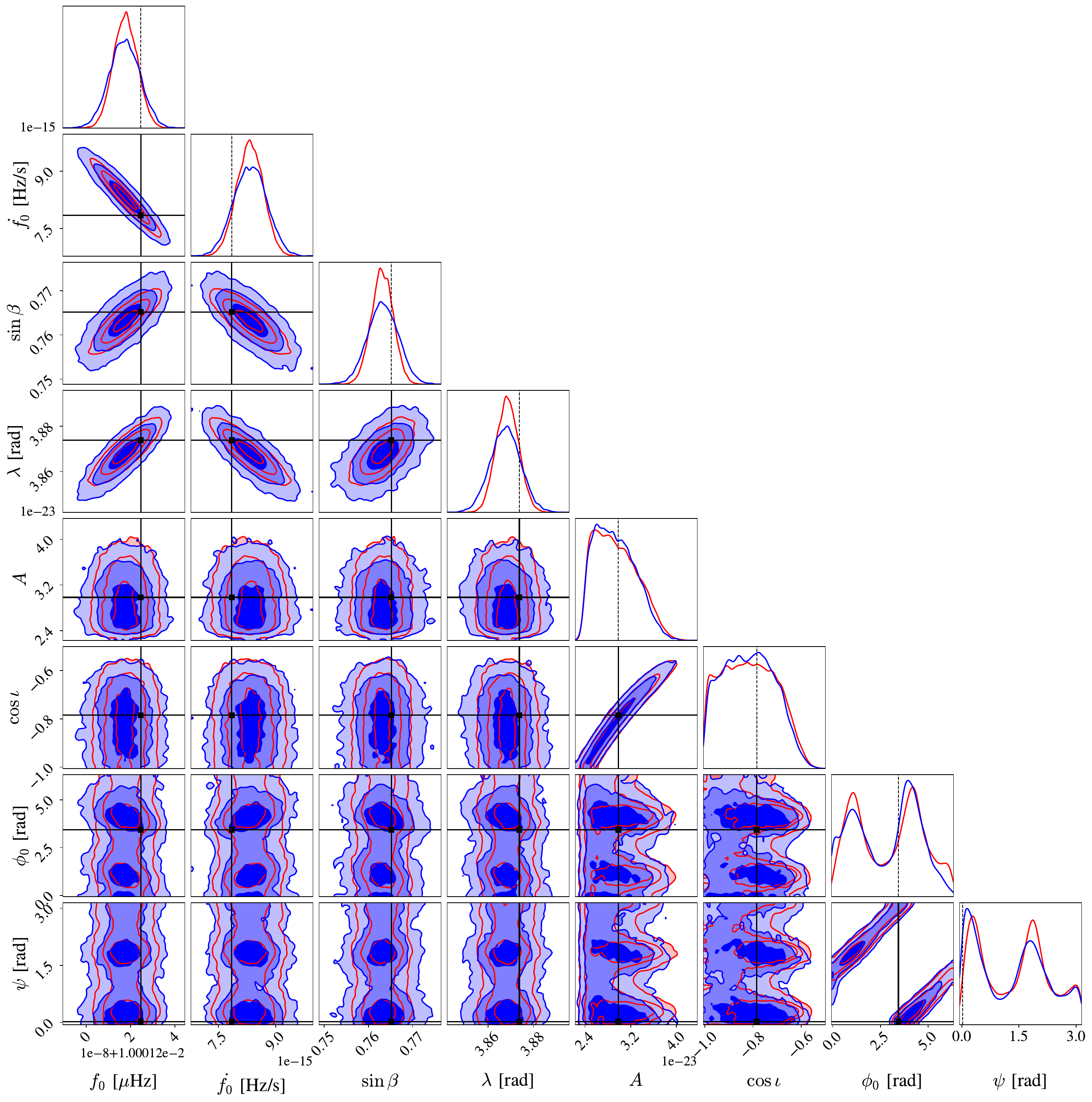}
\caption{
High-frequency case: empirical one- and two-dimensional marginal posterior distributions obtained by the proposed neural posterior estimator (blue) and the MCMC sampling method (red). The contours correspond to the 1$\sigma$, 2$\sigma$, and 3$\sigma$ credible regions, while the dark lines indicate the true parameter values.}
\label{fig:10mHz}
\end{figure*}

We present in Fig.~\ref{fig:10mHz} the corner plot comparing the empirical one- and two-dimensional posterior distributions estimated by the proposed neural posterior model and by the MCMC sampling algorithm. We observe that the model successfully reproduces the overall structure of the posterior distributions. However, the  posterior distributions recovered by the proposed approach are systematically broader than the ones estimated by the MCMC algorithm for the parameters $ f_{0}$, $\dot{f}_{0}$, $\lambda$ and  $\beta$. This difference can be understood by inspecting the waveform model given by \eqref{A_E_JKS}. The parameters $\left( A, \iota, \phi_{0}, \psi \right)$ primarily affect the amplitude of the waveform, resulting in a relatively smooth likelihood surface and, consequently, well-behaved posterior distributions that are easier to explore. In contrast, the parameters $\left( f_{0}, \dot{f}_{0}, \lambda, \beta \right)$ mainly impact the phase evolution of the waveform, leading to more complex posterior structures that are harder to explore.

Despite the aforementioned limitation, the proposed neural posterior estimator remains able to localize the source within the correct region of the parameter space and preserves the main correlations between parameters. However, the increased posterior variance with respect to the MCMC-based approach suggests that additional model capacity or further training improvements may be required in this regime. For instance, one solution would consist in granting the head of each ResNet with  an embedding network able to learn a compact summary statistics of the raw frequency-domain data $\tilde{A}(f)$ and $\tilde{E}(f)$ \cite{Dax2021a, AliceSparado}. Such a strategy may reduce the dimensionality of the ResNet input while preserving the information relevant for parameter inference. It  would also alleviate the need of increasing the number of network parameters as the number of frequency bins grows with the frequency.
Independently of this improvement, once trained, the neural posterior model can  serve as an informed proposal distribution to be embedded within RJMCMC samplers, which are one of the most computationally demanding components of the LISA global fit. Using the samples produced by the proposed neural posterior estimator to initialize new sources or design efficient jump proposals could significantly speed up the convergence of the sampler.

\subsection{Quantitative assessment of estimation performance}\label{subsec:JS}

\begin{figure}
\centering
\includegraphics[width=\columnwidth]{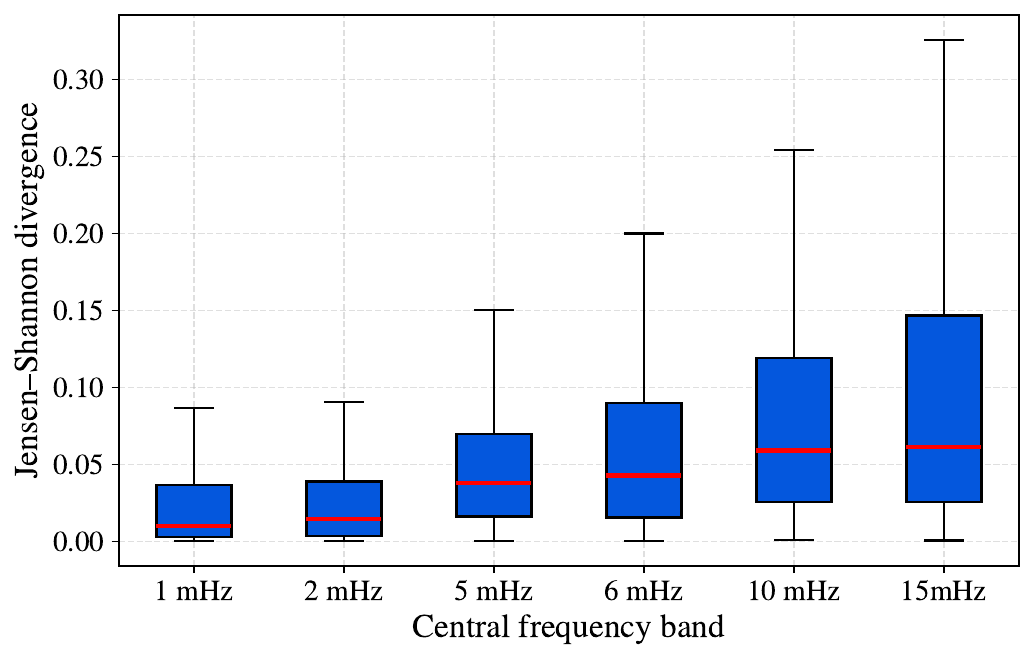}
\caption{
Performance as a function of the central frequency: box plots of the Jensen-Shannon divergence computed across parameters for each  neural posterior model trained at different central frequencies $f_{\mathrm{c}}$. Each box summarizes the variability of the JS divergences averaged over 100 realizations. The
lower and upper end of boxes represent the 25th and 75th percentile credible regions respectively. The lower and upper end
of the whiskers represent the 5th and 95th percentile credible regions. The middle lines are representative of the median JS
values.
}
\label{fig:js}
\end{figure}

So far, the comparison between the neural posterior estimation and the MCMC method has been carried out by visually inspecting a limited number of corner plots each depicting the  posterior distributions associated with a single waveform. To quantitatively support this comparison, we propose to compute the Jensen-Shannon (JS) divergence between the marginal posterior distributions recovered by the two approaches when analyzing a set of 100 waveforms. The JS divergence is a symmetric and bounded measure of discrepancy between two distributions $P$ and $Q$ and  defined as \cite{Lin1991a}
\begin{equation}
\mathcal{D}_{\mathrm{JS}}(P \,\|\, Q) = \frac{1}{2} \mathcal{D}_{\mathrm{KL}}(P \,\|\, M) + \frac{1}{2} \mathcal{D}_{\mathrm{KL}}(Q \,\|\, M)\,,
\label{DJS}
\end{equation}
with $M = \frac{1}{2}(P + Q)$. It takes its values in the interval $[0, 1]$, where $0$ indicates identical distributions and larger values correspond to increasing dissimilarity. Fig.~\ref{fig:js} depicts the average JS divergences as box-plots across the considered  values of the  central frequency.  We first observe that the JS divergence increases with the frequency. This trend is related to the increasing complexity of the signal as the frequency rises. In the low-frequency regime, the median JS divergence remains below $5 \times 10^{-2}$, indicating excellent agreement between the  neural posterior estimator and the  MCMC-based estimator. In the high-frequency regime, the median JS divergence is higher than $5 \times 10^{-2}$ but remains below $10^{-1}$, indicating an overall good agreement, yet with noticeable differences and broader posterior distributions.

\section{Exploring more realistic scenarios}
\label{sec:Extension}

Here we propose to explore challenging yet more realistic experimental scenarios. We first try to extend the previous analysis to wider frequency ranges in subsection~\ref{sec:multi_freq_band} by adding a frequency information to the flow. Then subsection \ref{sec:two_sources} explores the case of two overlapping GB sources.

\subsection{Multi frequency bands}
\label{sec:multi_freq_band}

In section \ref{sec:narrow_freq}, we trained neural posterior models on  very narrow frequency bands of $4\,\mu\mathrm{Hz}$, since the prior on $f_0$ can be highly constrained. However, scaling this approach to the full LISA frequency range would require training thousands of frequency-specific models, which is not computationally feasible. As an alternative, we propose hereafter to extend the width of the analysis window to $0.5\,\mathrm{mHz}$. This window width is chosen because the signal morphology remains approximately unchanged over this frequency range, and because covering the full LISA band would then require only a few tens of independently trained networks, which is computationally more tractable.\\

\begin{table}
\caption{\label{tab:waveform_params_mutli}
Multi frequency case: true waveform parameters.}
\begin{ruledtabular}
\begin{tabular}{lclc}
\textbf{Parameters} & & \textbf{Parameters} & \\
\hline
$f_0 \; [\mathrm{mHz}]$                 & $4.1317$             & $A$                        & $8.59 \times 10^{-23}$ \\
$\dot{f}_{0} \; [\mathrm{Hz\,s^{-1}}]$ & $4.41 \times 10^{-16}$ & $\iota \; [\mathrm{rad}]$  & $1.17$ \\
$\beta \; [\mathrm{rad}]$               & $0.56$                & $\phi_0 \; [\mathrm{rad}]$ & $3.88$ \\
$\lambda \; [\mathrm{rad}]$             & $1.13$                & $\psi \; [\mathrm{rad}]$   & $0.99$ \\
\end{tabular}
\end{ruledtabular}
\end{table}

\begin{figure*}
\centering
\includegraphics[width=\textwidth]{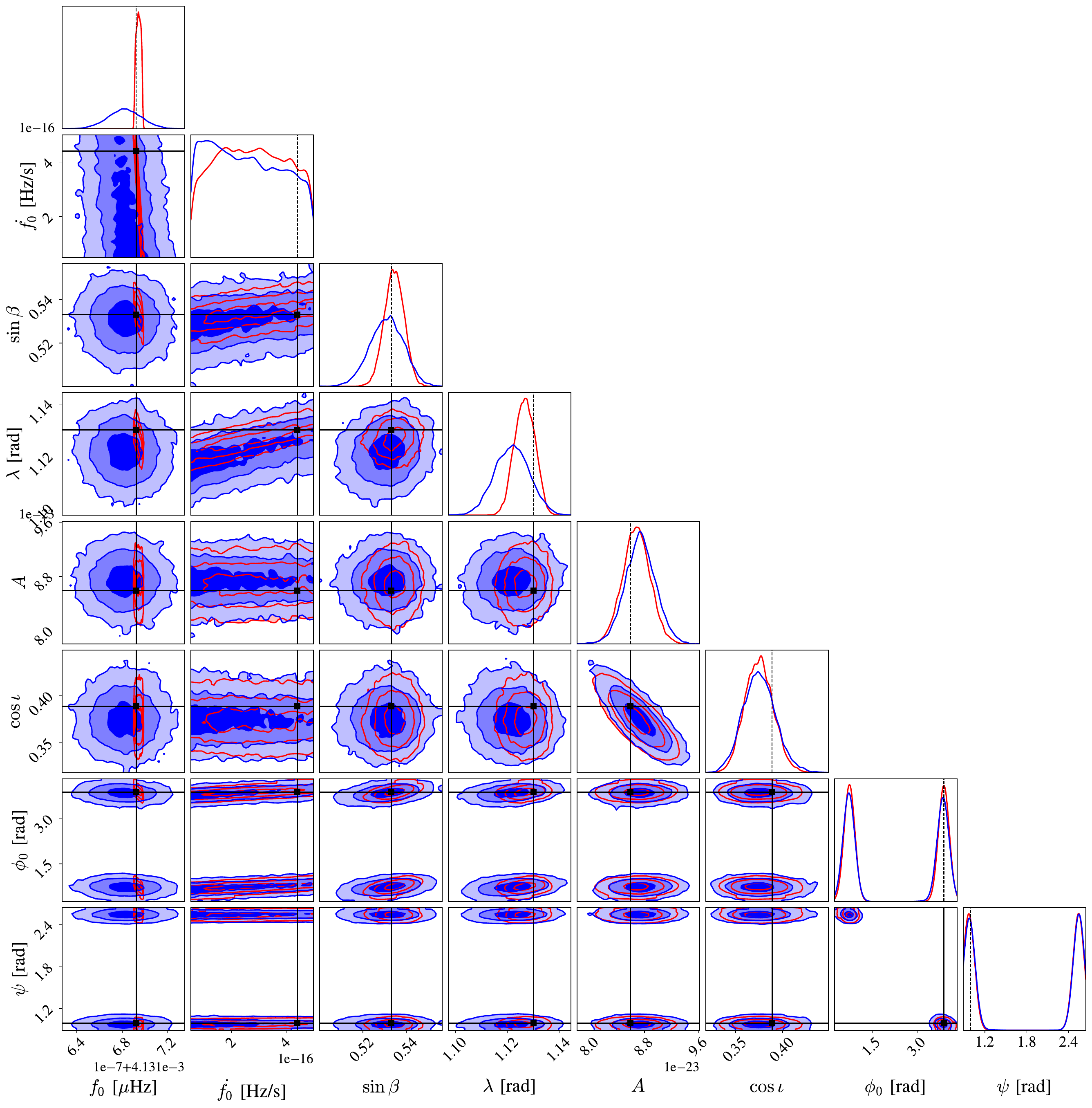}
\caption{
Multi frequency case: empirical one- and two-dimensional marginal posterior distributions obtained by the proposed neural posterior estimator (blue) and the MCMC sampling method (red). The contours correspond to the 1$\sigma$, 2$\sigma$, and 3$\sigma$ credible regions, while the dark lines indicate the true parameter values.}
\label{fig:multi_freq}
\end{figure*}

\noindent \textbf{Experimental protocol --} We consider the same setup as before, i.e., a very narrow frequency band of $4\,\mu\mathrm{Hz}$ containing a waveform. However, we now provide the network with additional frequency information, namely the central frequency $f_{\mathrm{c}}$ of the window. With this setup, the neural posterior estimator can be provided with a waveform at any frequency while retaining information about its location within the LISA spectrum through the central frequency of the corresponding analysis window. In practice, this is implemented by concatenating the input signal $d$  with the central frequency $f_{\mathrm{c}}$ of the window.

The dataset is constructed as before, based on the same priors reported in Table~\ref{tab:priors}, except for the frequency $f_0$. We first randomly select a central frequency $f_{\mathrm{c}}$ in a $0.5\,\mathrm{mHz}$-width spectral range arbitrarily fixed as $\left[4.0\,\mathrm{mHz}, 4.5\,\mathrm{mHz}\right]$, i.e., $f_{\mathrm{c}} \sim \mathcal{U}\left[4.0\,\mathrm{mHz}, 4.5\,\mathrm{mHz}\right]$. We then set a frequency window of width $4\,\mu\mathrm{Hz}$ around $f_{\mathrm{c}}$, as in the previous section. Next, we draw a frequency $f_0$ within this window, i.e., $f_0 \sim \mathcal{U}\left[f_{\mathrm{c}} - 2\,\mu\mathrm{Hz}, f_{\mathrm{c}} + 2\,\mu\mathrm{Hz}\right]$, and then construct the waveform as shown in Fig.~\ref{fig:gb}. The neural posterior estimator is provided with the concatenation of the waveform $d$ and the central frequency $f_{\mathrm{c}}$. The network architecture and hyperparameters are the same as in the low-frequency case, resulting in approximately $1.2 \times 10^8$ trainable parameters.\\

\noindent \textbf{Results --} Fig.~\ref{fig:multi_freq} compares the samples generated by the trained neural posterior model with those obtained from MCMC sampling, based on the true waveform parameters reported in Table~\ref{tab:waveform_params_mutli}. We observe that there is again very good agreement between the posteriors of the parameters $\left( A, \iota, \phi_{0}, \psi \right)$. This can be explained by the fact that the shape of the posterior distributions for these parameters does not depend on the frequency. Indeed, as discussed in the previous section, these four particular parameters define the amplitude of the signal and therefore affect the waveform in the same way regardless of the frequency. In contrast, the posterior distributions of the parameters $\left( \beta, \lambda \right)$ vary with frequency. Consequently, the neural posterior model faces a more challenging learning task and must capture a more complex mapping for these parameters, as their behavior varies across frequencies. 

\begin{figure}
\centering
\includegraphics[width=\columnwidth]{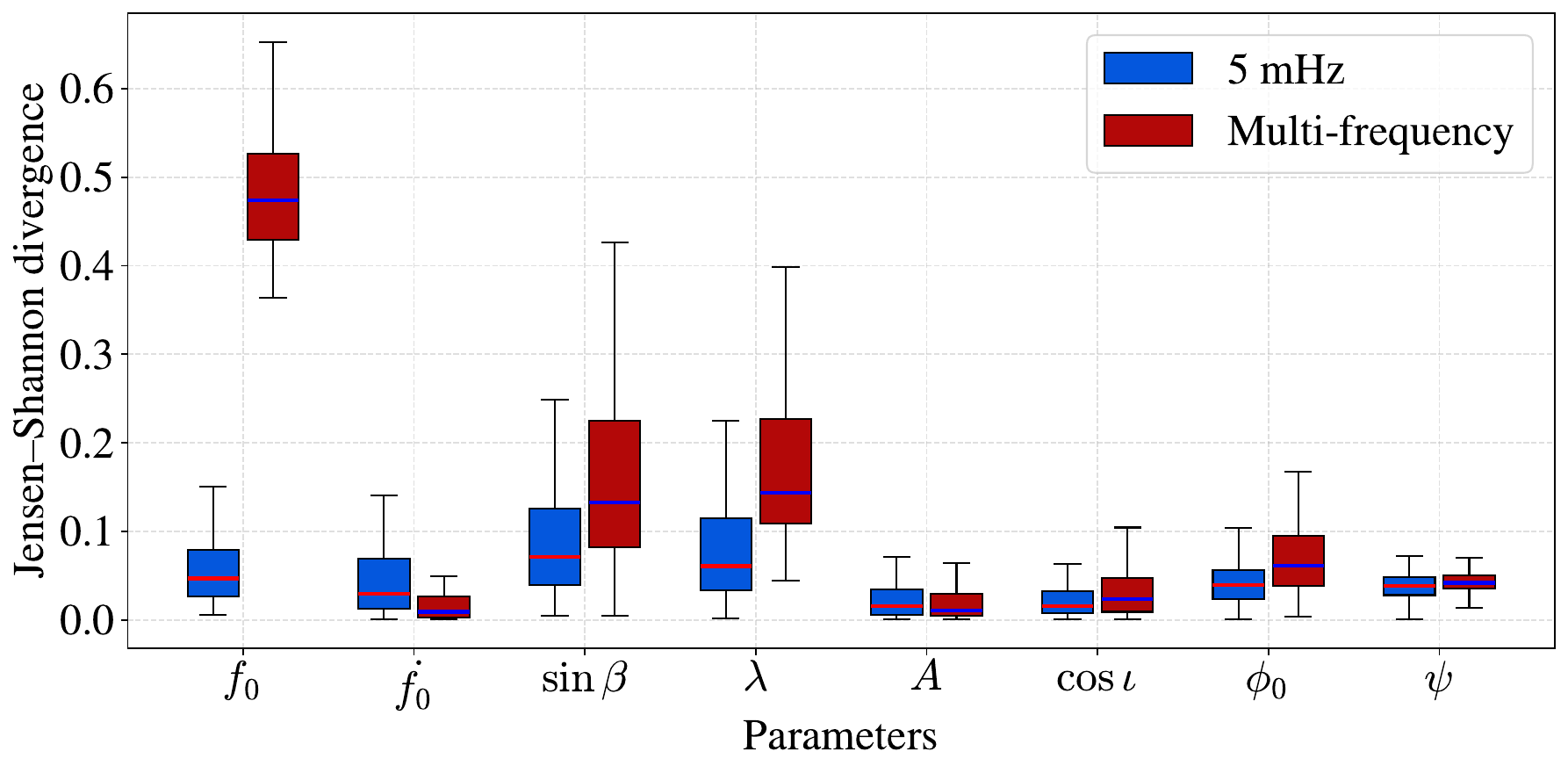}
\caption{\label{fig:js_multi_freq}
Multi frequency case: box plots of the Jensen--Shannon divergence between distributions computed across parameters for the narrow frequency case at 5 mHz and the multi frequency band case. Each box summarizes the variability of the JS divergences averaged over 100 realizations. The lower and upper end of boxes represent the 25th and 75th percentile credible regions respectively. The lower and upper end
of the whiskers represent the 5th and 95th percentile credible regions. The middle lines are representative of the median JS
values.
}
\end{figure}

To quantitatively support this findings, as in subsection \ref{subsec:JS}, the discrepancy between the marginal posterior distributions recovered by MCMC sampling and the neural posterior estimator is assessed by averaging their JS divergence over 100 test waveforms. More specifically, Fig.~\ref{fig:js_multi_freq} reports the results obtained in the considered multi-frequency bands as well as those  obtained in the case of a $5$ mHz narrow band. These results confirm the conclusions drawn from Fig.~\ref{fig:multi_freq}. In both cases, the compared methods lead to close estimations for parameters $\left( A, \iota, \phi_{0}, \psi \right)$. On the contrary, the parameters associated with the sky localization exhibit a wider dispersion and a higher median value in the case of multi-frequency band than in the case of a narrow band. Finally, while the parameter $f_0$ was well estimated in the the narrow-frequency case, the multi-frequency case appears to lead to a systematic broader posterior.

Finally, from the results depicted in Fig.~\ref{fig:multi_freq} and Fig.~\ref{fig:js_multi_freq}, the frequency $f_0$ appears to be substantially more difficult to estimate than in the previous experimental scenario, due to the much broader spectral support of its prior distribution. Indeed, the ratio between the prior width and the posterior uncertainty is significantly larger than in the previous analysis: the prior width increases from $4\,\mu\mathrm{Hz}$ in the previous analysis to $0.5\,\mathrm{mHz}$, corresponding to a parameter space that is more than two orders of magnitude larger. Furthermore, the posterior distribution of the parameter $f_0$ is expected to be concentrated within a extremely narrow support, with a width of $\sim10^{-3}\,\mu\mathrm{Hz}$. As the neural posterior estimator must resolve a highly localized posterior region within a broad prior support, accurate coverage of the parameter space may require extending both the training dataset and the training duration. Another possible strategy would be to introduce an embedding network that maps the waveform and central frequency into a more expressive representation before feeding them to the neural posterior estimator.

\subsection{Two sources}
\label{sec:two_sources}

In the previous sections, the performance of the proposed neural posterior estimator was assessed while analyzing a single GB. However, LISA is expected to detect thousands of GBs overlapping in both the time and frequency domains \cite{GBs}. Developing methods capable of correctly inferring posterior distributions in the presence of overlapping sources is therefore essential for analyzing the LISA strain data. To match these demanding requirements, we extend our analysis to the case of two overlapping sources and evaluate how the proposed neural posterior model is able to learn posterior distributions in this higher-dimensional setting.\\

\begin{table}
\caption{\label{tab:2_sources}
Two-source case: true waveform parameters.}
\begin{ruledtabular}
\begin{tabular}{lcc}
 \textbf{Parameters} & \textbf{Favorable case} & \textbf{Unfavorable case} \\
\hline
$f_{01} \; [\mathrm{mHz}]$        & $5.0003$ & $5.0006$  \\
$\dot{f}_{01} \; [\mathrm{Hz\,s^{-1}}]$& $5.00 \times10^{-16}$ & $4.26 \times10^{-16}$  \\
$\beta_1 \; [\mathrm{rad}]$     & $0.30$ & $-0.21$  \\
$\lambda_1 \; [\mathrm{rad}]$   & $4.50$ & $3.31$  \\
$A_1$                           & $4.02 \times10^{-23}$ & $6.76 \times10^{-23}$ \\
$\iota_1 \; [\mathrm{rad}]$     & $0.80$ & $1.87$  \\
$\phi_{01} \; [\mathrm{rad}]$    & $1.31$ & $4.10$  \\
$\psi_1 \; [\mathrm{rad}]$      & $0.51$ & $0.37$  \\
$f_{02} \; [\mathrm{mHz}]$        & $5.0008$ & $5.0008$  \\
$\dot{f}_{02} \; [\mathrm{Hz\,s^{-1}}]$& $7.00 \times10^{-16}$ & $1.11 \times10^{-15}$  \\
$\beta_2 \; [\mathrm{rad}]$     & $0.01$ & $0.41$  \\
$\lambda_2 \; [\mathrm{rad}]$   & $2.20$ & $5.58$  \\
$A_2$                           & $2.06 \times10^{-23}$ & $4.30 \times10^{-23}$ \\
$\iota_2 \; [\mathrm{rad}]$     & $0.70$ & $0.43$  \\
$\phi_{02} \; [\mathrm{rad}]$    & $0.82$ & $3.51$  \\
$\psi_2 \; [\mathrm{rad}]$      & $1.00$ & $2.51$  \\
\end{tabular}

\end{ruledtabular}
\end{table}

\noindent \textbf{Experimental protocol --} We consider  the same setup as in section ~\ref{sec:narrow_freq}, with a narrow frequency band around a central frequency $f_{\mathrm{c}}=5$ mHz. The training dataset is the same as in the single-source case. During training, we draw two sets of quadruplet pairs $(\tilde{A}_{\mu}^{(n)}, \tilde{E}_{\mu}^{(n)})$ from their predefined set, each pair being defined by the parameters $\left( f_{0}, \dot{f}_{0}, \lambda, \beta \right)$  (see \eqref{A_E_JKS}) that are denoted as $\left( f_{01}, \dot{f}_{01}, \lambda_1, \beta_1 \right)$ and $\left( f_{02}, \dot{f}_{02}, \lambda_2, \beta_2 \right)$, respectively. Then two sets $\left( A_1, \iota_1, \phi_{01}, \psi_1 \right)$ and $\left( A_2, \iota_2, \phi_{02}, \psi_2 \right)$ of the remaining parameters are randomly drawn from their prior distributions. The two waveforms are computed according to \eqref{A_E_JKS}. We ensure that the SNR of each source is in the range  10-100, otherwise we draw new samples of the parameters $\left( A, \iota, \phi_{0}, \psi \right)$. Finally, the two waveforms are summed before  feeding the resulting signal into the neural posterior model. Preliminary results obtained from this experimental setup showed the presence of two modes in the posterior distributions predicted by the neural posterior model, each corresponding to one of the two sources. To avoid identifiability issues, the order of the parameters are thus enforced  during training by systematically assigning the smaller value of the frequency parameter $f_0$ to the first source (i.e., $f_{01} \leq f_{02}$, by construction). The network architecture and hyperparameters are the same as in the low-frequency case, except that the number of flow steps has been increased to 18 to account for the higher dimensionality of the problem. This results in approximately $1.7 \times 10^8$ trainable parameters.\\

\begin{figure*}
\centering
\includegraphics[width=\textwidth]{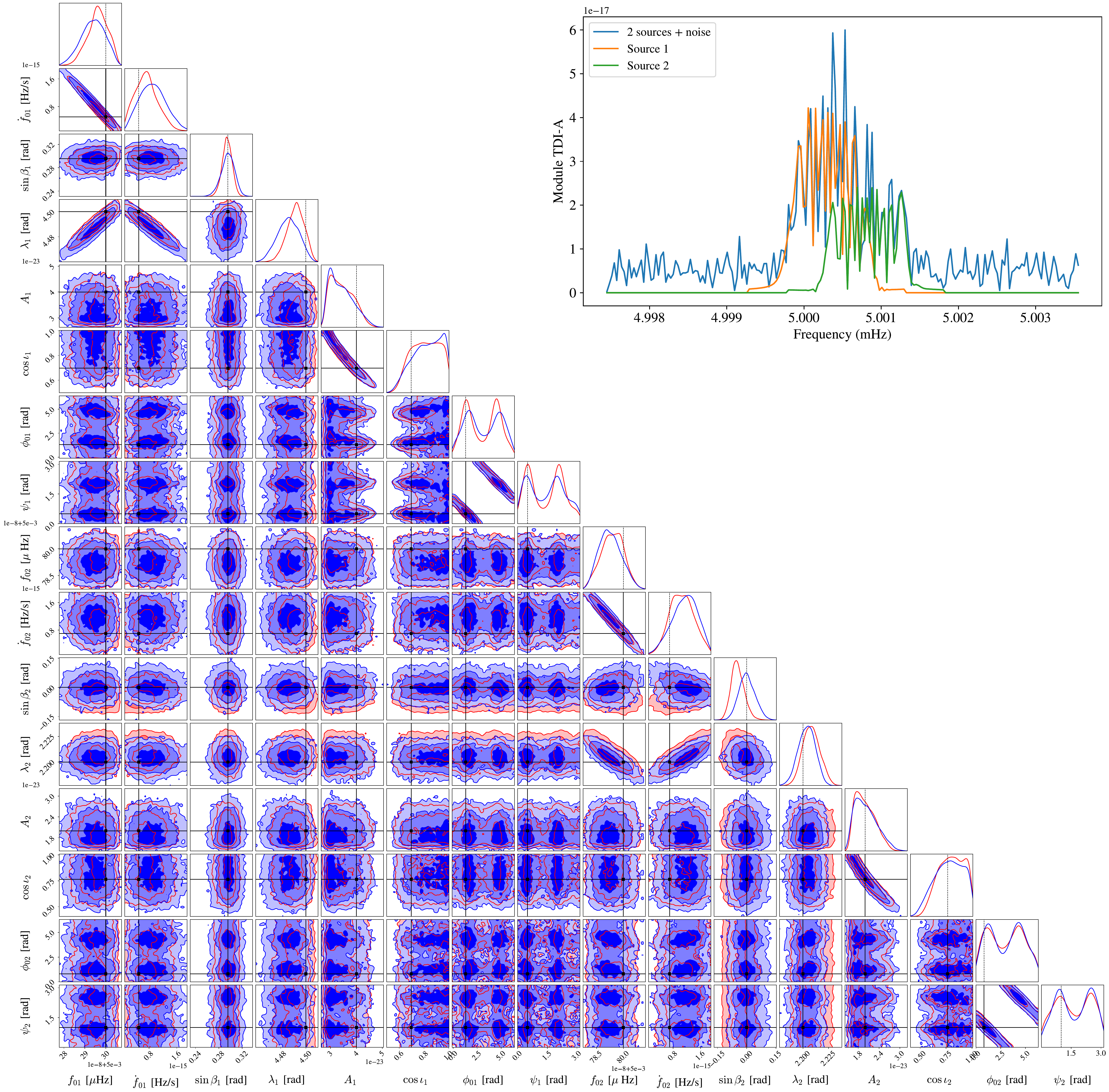}
\caption{
Favorable two-source case: empirical one- and two-dimensional marginal posterior distributions obtained by the proposed neural posterior estimator (blue) and the MCMC sampling method (red). The contours correspond to the 1$\sigma$, 2$\sigma$, and 3$\sigma$ credible regions, while the dark lines indicate the true parameter values. We add the plot of the data for the module of TDI-A in blue and the two injected waveform in orange and green.
}
\label{fig:2_sources}
\end{figure*}

\begin{figure*}
\centering
\includegraphics[width=\textwidth]{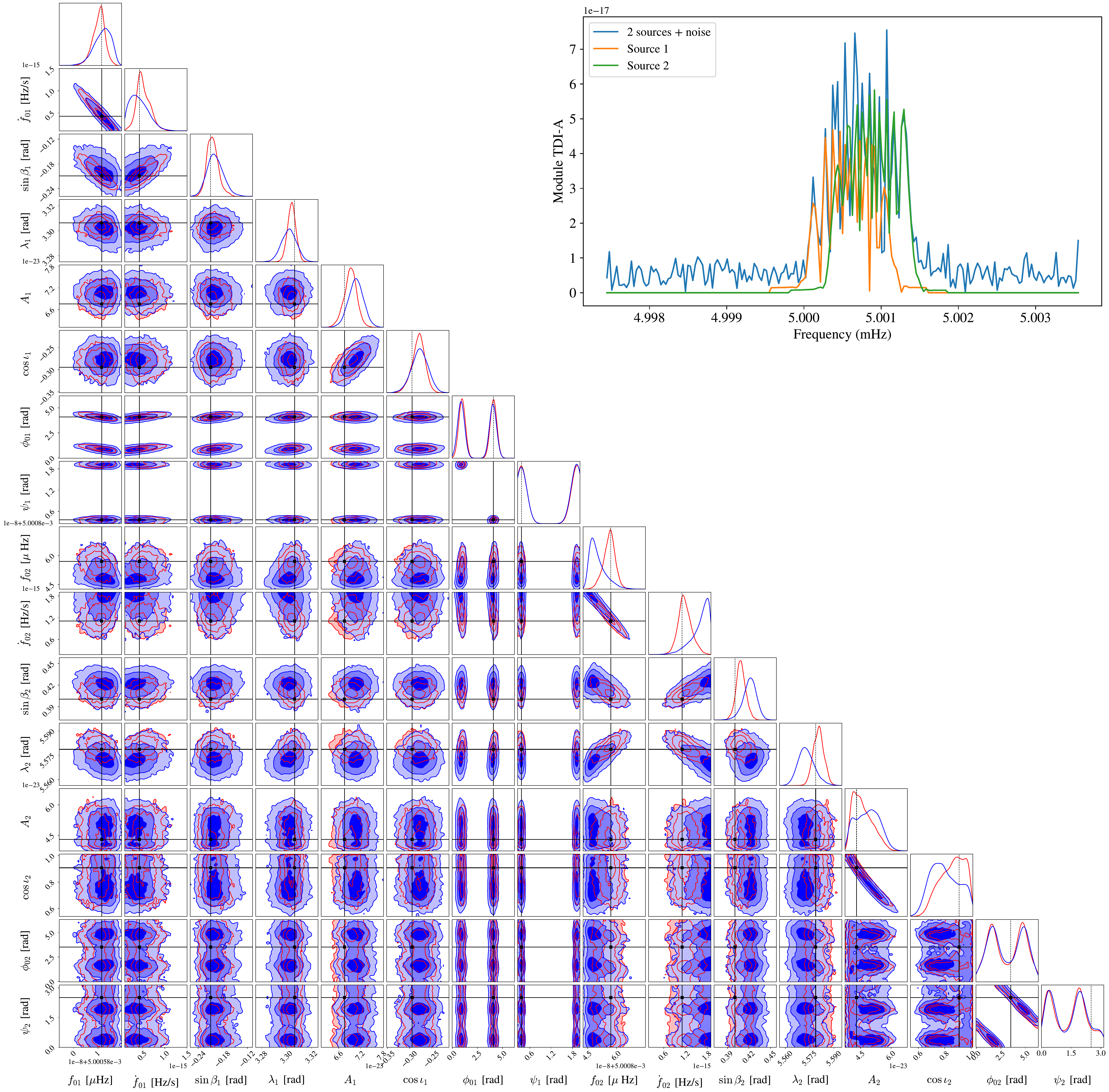}
\caption{
Unfavorable two-source case: empirical one- and two-dimensional marginal posterior distributions obtained by the proposed neural posterior estimator (blue) and the MCMC sampling method (red). The contours correspond to the 1$\sigma$, 2$\sigma$, and 3$\sigma$ credible regions, while the dark lines indicate the true parameter values. We add the plot of the data for the module of TDI-A in blue and the two injected waveform in orange and green.}
\label{fig:2_sources_bad}
\end{figure*}

\noindent \textbf{Results --} In contrast to the previous results, we observe a large diversity of results in the experimental scenario of two overlapping sources. Some marginal posteriors inferred by the neural posterior model are very close to those obtained with MCMC sampling, while others are overly broad. As typical example results, we considered two different cases, one which is favorable while the other is less favorable, each characterized by true parameters  reported in Table~\ref{tab:2_sources}. The estimated one- and two-dimensional marginal posterior distributions are depicted in Fig.~\ref{fig:2_sources} and~\ref{fig:2_sources_bad}, respectively. On their top-right corner, each figure also represents  the module of the input TDI A response as well as the two waveforms to appreciate how they overlap. 

In Fig.~\ref{fig:2_sources}, corresponding to the favorable case, we observe generally good agreement between the posterior distributions recovered by the proposed neural posterior estimator and by MCMC sampling. For some parameters, these results are comparable to those obtained in the low-frequency case discussed in Section~\ref{sec:low_frequency_case}. In particular, the complex and multi-modal shapes of the posterior distributions are successfully captured by the proposed estimator.  For other parameters, however, the posteriors are slightly broader than those obtained in the best previous results; however, their overall structures are well recovered. 

Conversely, in Fig.~\ref{fig:2_sources_bad}, corresponding to the unfavorable case,  we observe that  the distributions recovered by the neural posterior estimator are much broader for most parameters than those obtained with MCMC sampling. For some parameters, the neural posterior model is unable to properly recover the shape of the posterior distribution or provides significantly biased estimates.

The differences observed across the more and less favorable cases can be attributed to the extent of overlap between the two waveforms, which is partly driven by the respective frequencies associated to the two GB sources. As specified in Table~\ref{tab:2_sources}, the frequency separation between the two sources is  $\Delta f_0 = \left| f_{01} - f_{02} \right| = 5 \times 10^{-4}$ mHz in the favorable case and $\Delta f_0 = 2 \times 10^{-4}$ in the unfavorable case. In other words, the favorable configuration corresponds  to  two partially overlapping waveforms whereas the unfavorable configuration involves   two nearly fully overlapping waveforms, making the parameter estimation task significantly more difficult. 

\begin{figure}
\centering
\includegraphics[width=\columnwidth]{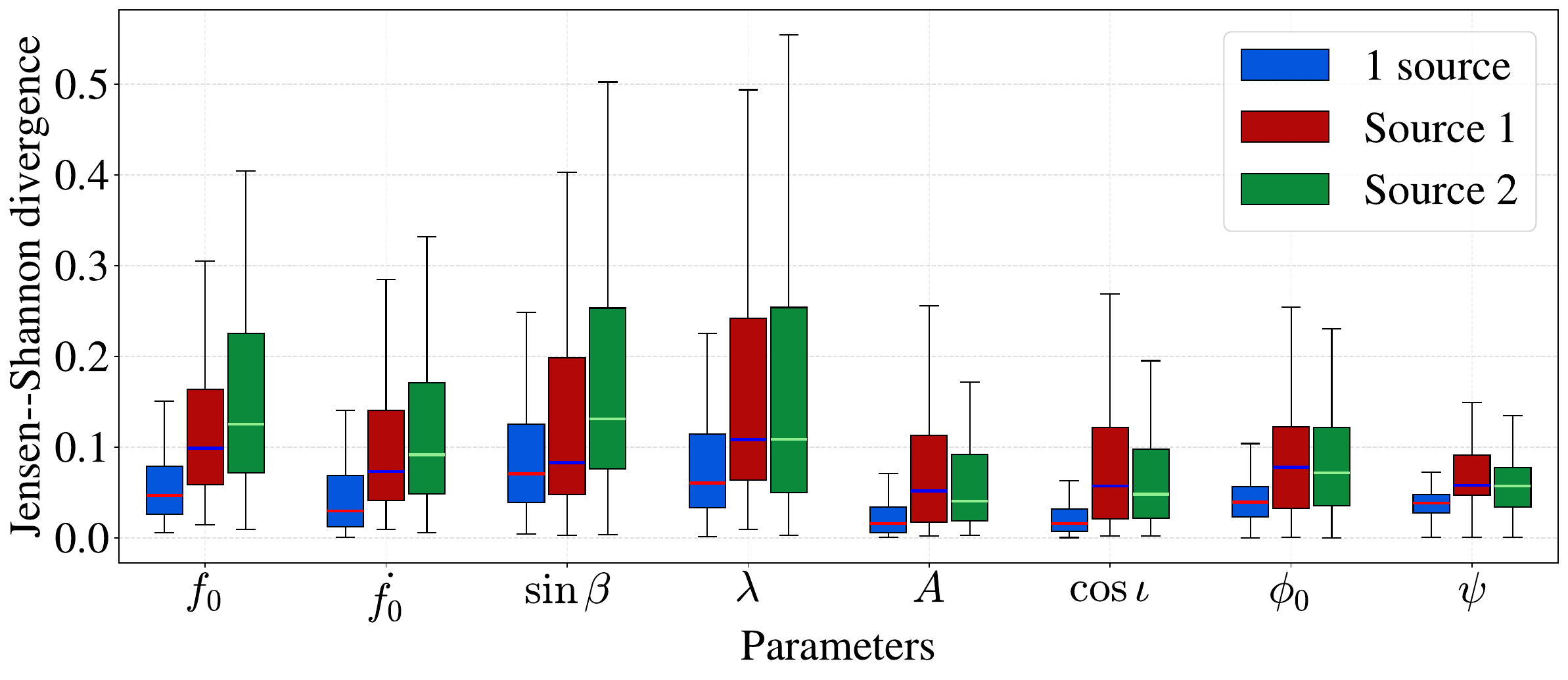}
\caption{\label{fig:js_2_sources}
Two-source case: box plots of the Jensen--Shannon divergence computed across parameters for the narrow frequency case at 5 mHz and the two-source case. Each box summarizes the variability of the JS divergence averaged over 100 realizations. The lower and upper end of boxes represent the 25th and 75th percentile credible regions respectively. The lower and upper end of the whiskers represent the 5th and 95th percentile credible regions. The middle lines are representative of the median JS
values.
}
\end{figure}

To quantitatively support this experimental analysis, the discrepancy between the recovered posterior distributions is evaluated by computing the JS divergence averaged over 100 test signals. Fig.~\ref{fig:js_2_sources} compares the results obtained in the considered two-source case with those individually obtained in the $5$ mHz narrow-band case. We observe that for all parameters, the JS divergences exhibit a wider range of possible values and higher median values than in the single-source case. This means that a large diversity of results can be obtained, with some signals leading to good agreement between the posteriors recovered by the neural posterior estimator and MCMC sampling but some signal leading to significant differences between the estimated posterior distributions. This shows that analyzing signals composed of two overlapping sources is more challenging  than analyzing a single source. Yet, again,  it is worth noting that the posterior samples produced by proposed neural posterior model can serve as informed proposals or initializations to be included into RJMCMC algorithms, which could significantly accelerate the convergence of the samplers.

\section{Conclusions and future works}
\label{sec:conclusion}

In this work, we have explored the use of NPE via NF for the Bayesian inference of GB parameters as observed by LISA. Our results demonstrate that SBI offers a compelling alternative to classical MCMC-based methods, providing thousands of independent posterior samples in less than a second once the network is trained, without any likelihood evaluation.

For single-source parameter estimation in narrow frequency bands, the method performs very well in the low-frequency regime (1-6\,mHz), as confirmed both visually through corner plots and quantitatively through p-p plots and JS divergence measures. In the high-frequency regime (7-15\,mHz), performance degrades somewhat, which we attribute to the increased complexity of the signal morphology and the larger number of frequency bins required to represent the waveform. Importantly, at this stage of the work we deliberately chose not to use an embedding network for the input data, in order to characterize the performance of the flow in isolation and avoid conflating potential errors introduced by the embedding with those of the flow itself. The introduction of a dedicated embedding network is expected to significantly improve performance in the high-frequency regime by compressing the input into a more informative and manageable representation. Even in its current form, however, the flow-based posteriors are sufficiently accurate to serve as informative proposals within classical MCMC and RJMCMC samplers, a concrete and near-term benefit for global fit pipelines.

Towards more realistic scenarios, we have also investigated two extensions of the single-source case. First, we extended the analysis to a wider frequency band of $0.5\,\mathrm{mHz}$, by informing the network with the central frequency of the analysis window. Results on the amplitude parameters remain comparable to the narrow-band case, while sky localization parameters prove slightly harder to recover, and $f_0$ in particular becomes significantly more challenging to estimate due to the broader prior range. This approach is a first step toward covering the full LISA band with a tractable number of networks, and informing the network with the frequency via embeddings will be investigated in future work as a way to address the remaining limitations. Second, we demonstrated that the method can be extended to the two-source case, where two GBs overlap within the same frequency window. Unlike the single-source case, results exhibit a significant diversity across configurations: some posteriors are in very good agreement with those provided by MCMC sampling, while others are broader, reflecting the increased complexity of the two-source inference problem. This was expected given the increased dimensionality and the difficulty of disentangling two superimposed waveforms, and motivates further developments in network capacity and training strategies.

Several important directions remain open for future work. Scaling the approach to cover the full LISA frequency band will require determining the optimal number and placement of analysis windows, and embedding networks will play a central role in making this feasible. On the source-number side, the present work assumes a fixed and known number of sources, while in reality the number of signals in any given frequency band is unknown. Addressing this trans-dimensional problem is a key next step. One natural way is to investigate the robustness of a fixed-dimension network when presented with data containing a different number of sources than assumed during training. Beyond that, one can envision incorporating a classification-like mechanism to switch between network regimes depending on the estimated number of sources, or training the network directly on outputs of variable dimension. Finally, one natural long-term perspective would be a fully SBI-based approach to GB inference at the scale of the whole Galaxy. In the meantime, even imperfect posteriors from the current framework can already provide immediate practical value: the flow-based posteriors can be used as informative proposals within classical MCMC and RJMCMC global fit pipelines, replacing generic proposals with data-driven density estimates derived directly from the observations. Furthermore, the posteriors produced by the current network can serve to initialize new sources or to seed sampler chains in a region of high posterior mass. This has the potential to significantly accelerate convergence and improve the overall performance of global fit analyses such as those currently being developed for LISA.

A further challenge that will need to be addressed concerns the realism of the noise model assumed during training. In the present work, we adopt a fixed, stationary noise PSD, which provides a clean and controlled setting for evaluating the performance of the flow-based approach, but which does not reflect the complexity of actual LISA data. In practice, the noise PSD will vary over time due to instrumental drifts, and the data stream will be affected by non-stationary artifacts such as glitches and gaps arising from scheduled or unscheduled interruptions of the instrument. These features are particularly challenging for GB inference: gaps break the phase coherence of the quasi-monochromatic signal and make the Fourier domain less ideal as a data representation, while glitches can mimic or mask source signals, and a mismatched noise model can introduce systematic biases in the recovered posteriors. In global-fit analyses, this issue is further compounded by the fact that the noise PSD cannot be estimated independently of the GW signals and must instead be inferred jointly; the PSD estimated at a previous stage of the analysis could therefore be used to the network with inference time, ensuring consistency between source parameter estimation and the evolving noise model. One of the key advantages of the SBI framework is precisely that such realistic data features can in principle be incorporated into the training data without requiring any modification to the inference procedure itself, since no explicit likelihood is evaluated at inference time. This stands in contrast to classical MCMC approaches, where handling non-stationary noise or data gaps typically requires dedicated pre-processing, inpainting, or non-diagonal noise covariance matrix formulations, each introducing additional approximations or computational overhead. Training the model on a representative ensemble of noise realizations that includes variable PSDs, glitch artifacts, and gap patterns possibly in a time-frequency representation instead of a Fourier-domain representation would therefore be a natural and principled extension of the present framework towards operational LISA data analysis conditions.

\begin{acknowledgments}
We would like to acknowledge CNES for funding this project and this PhD thesis within the LISA framework. N.K. would like to thank Stas Babak for useful discussions at the beginning of the project especially regarding JSK parameterization. N.K. gratefully acknowledges support from the CNES for the exploration of LISA science and support from the CNRS/IN2P3 Computing Center (Lyon - France) for providing computing and data-processing resources needed for this work. N.D. and T.O. acknowledge partial support from the Artificial and Natural Intelligence Toulouse Institute (ANITI), funded by the France 2030 program under the grant agreement ANR-23-IACL-0002.  
\end{acknowledgments}

\section*{Appendix}
\appendix

\subsection{LISA response}
\label{app:LISAResponse}

The GW tensor in equation \eqref{h_tensor} is projected on the three arms of LISA. We denote by $n_{ij}$ ($i,j \in \left\{ 1,2,3 \right\}$) the unit vector along the arm of LISA between the spacecrafts $i$ and $j$, and $L_{ij}$ the length of the arm between the spacecraft $i$ and $j$.

The LISA response \cite{Neil2003a, Neil2003b, LISA2} is constructed from single-link observables $y_{ij}$, which represent
the relative distance shift measured between the transmitting spacecraft
$i$ and the receiving spacecraft $j$. These observables are defined as
\begin{equation}
y_{ij}(t) = \frac{\delta L_{ij}(t)}{2L_{ij}}\,.
\label{yij}
\end{equation}
The distance shift measured is the projection of the GW along the detector arm
\begin{equation}
\delta L_{ij}(t) =
\frac{1}{2}
\frac{n_{ij}(t) \otimes n_{ij}(t)}{1 - k \cdot n_{ij}}
:
\int_{t_{i}}^{t_{r}}\boldsymbol{h}(t')dt'\,,
\label{dlij}
\end{equation}
with $t_i$ and $t_r$ the emission and reception times of the photon at spacecraft $i$ and $j$, respectively, and the notation $X:Y = X_{lm}Y_{lm}$.

Decomposing the gravitational wave tensor \eqref{h_tensor} on its frequency components defined as
\begin{equation}
    \textbf{h}(t) = \int_{-\infty}^{+\infty}\tilde{\textbf{h}}(f)e^{-2 \pi i f t}df\,,
\end{equation}
we can rewrite the equation \eqref{dlij} as \cite{Neil2003a}
\begin{equation}
\delta L_{ij}(t) = L_{ij}\int_{-\infty}^{+\infty}\textbf{R}(f, t, k):\tilde{\textbf{h}}(f)e^{2 \pi i f t}df\,,
\label{dlij_bis}
\end{equation}
with $\textbf{R}(f, t, k)$ the one-arm detector response tensor defined as
\begin{equation}
\textbf{R}(f, t, k) = \frac{1}{2}n_{ij}(t) \otimes n_{ij} (t)\mathcal{T}(f, t, k)\,,
\end{equation}
where the transfer function is given by \cite{Neil2003a}
\begin{equation}
\begin{aligned}
\mathcal{T}(f, t, k) =
& \ \mathrm{sinc}\Big( \frac{f}{2f^{*}_{ij}}(1-k\cdot n_{ij}(t)) \Big) \\
& \times \exp\Big( i \frac{f}{2f^{*}_{ij}}(1-k\cdot n_{ij}(t)) \Big)\,,
\end{aligned}
\end{equation}
with $f^{*}_{ij} = c/2\pi L_{ij}$ the transfer frequency for the $ij$-arm.

Equations \eqref{dlij} and \eqref{dlij_bis} correspond to the entire response of the $ij$-arm for a GW for all possible frequencies. However, for GBs, the time evolution of their frequency is extremely slow \eqref{phase_GB} compare to the travel time of light from a spacecraft $i$ to $j$, $f/\dot{f} \ll L/c$. This adiabatic condition implies that the frequency can be considered approximately constant over the photon propagation time inside the interferometer. Writing $t' = t_i + \delta t'$, we have \cite{Neil2003a}
\begin{equation}
    \Phi(t') = 2\pi(f_0 + \dot{f}_0t_i)\delta t' + \mathrm{const}\,.
\end{equation}
Thus, the GW can be treated as locally monochromatic during the integration, with a slowly varying frequency updated between successive intervals. The detector tensor $\mathbf{R}(f,t,k)$ can then be evaluated at a single representative frequency, and taken outside the integral. Taking the real part of the expression, the final LISA response for a GB is \cite{Neil2003a}

\begin{equation}
\delta L_{ij}(t)
= L_{ij}\,\Re\!\left[
\textbf{R}(f,t,k) : \textbf{h}(t)
\right]\,,
\label{LISA_response_GB}
\end{equation}
with $\textbf{h}(t)$ the complex form of the gravitational tensor \eqref{h_tensor}.

\subsection{Time-Delay Interferometry}
\label{app:TDI}

The single-link observables $y_{ij}$ \eqref{yij} are dominated by the laser frequency noise,
which is several orders of magnitude larger than the expected gravitational-wave
signal \cite{Massimo1999a}. To suppress this noise, LISA employs TDI \cite{Massimo1999a, Neil2003c, Massimo2004a, Katz2022b}. TDI consists in constructing specific linear
combinations of delayed single-link measurements such that the laser noise
cancels out.

In the first-generation TDI, three independent channels $X$, $Y$, and $Z$ are defined as combinations of the single-link measurements $y_{ij}$ \eqref{yij} with appropriate
time delays along the arms. In this approximation, the spacecraft follow Keplerian orbits, so the constellation is rotating around the sun. The arm lengths are assumed to be constant in time
but not equal to each other, i.e. $L_{12} \neq L_{23} \neq L_{31} \neq L_{21} \neq L_{13} \neq L_{32}$, which corresponds to a rigid but unequal-arm configuration. This differs from second-generation TDI, where the arm lengths vary with time. In this paper we used TDI first-generation.

We introduce the delay operator $\mathbf{D}$ as \cite{Katz2022b}
\begin{equation}
    \mathbf{D}_{i_{1},i_{2},...,i_{H}}x(t) = x\left( t-\sum_{h=1}^{H-1}L_{i_h i_{h+1}} \right)\,.
\end{equation}
We can then write the three TDI channels \cite{Katz2022b}
\begin{equation}
\begin{aligned}
X(t) &= y_{13}(t) + \mathbf{D}_{13}y_{31}(t)
+ \mathbf{D}_{131}y_{12}(t)
+ \mathbf{D}_{1312}y_{21}(t) \\
&\quad - \left[
y_{12}(t)
+ \mathbf{D}_{12}y_{21}(t)
+ \mathbf{D}_{121}y_{13}(t)
+ \mathbf{D}_{1213}y_{31}(t)
\right]\,, \\[6pt]
Y(t) &= y_{21}(t)
+ \mathbf{D}_{21}y_{12}(t)
+ \mathbf{D}_{212}y_{23}(t)
+ \mathbf{D}_{2123}y_{32}(t) \\
&\quad - \left[
y_{23}(t)
+ \mathbf{D}_{23}y_{32}(t)
+ \mathbf{D}_{232}y_{21}(t)
+ \mathbf{D}_{2321}y_{12}(t)
\right]\,, \\[6pt]
Z(t) &= y_{32}(t)
+ \mathbf{D}_{32}y_{23}(t)
+ \mathbf{D}_{323}y_{31}(t)
+ \mathbf{D}_{3231}y_{13}(t) \\
&\quad - \left[
y_{31}(t)
+ \mathbf{D}_{31}y_{13}(t)
+ \mathbf{D}_{313}y_{32}(t)
+ \mathbf{D}_{3132}y_{23}(t)
\right]\,.
\end{aligned}
\label{TDI}
\end{equation}
In the frequency domain, the TDI variables are obtained by taking the Fourier transform of the single-link measurements $y_{ij}$ \eqref{yij} and replacing each time delay $t - nL$ by a complex phase factor $e^{-2\pi i f nL}$. This follows directly from the Fourier shift theorem. Consequently, the Fourier-domain TDI combinations $\tilde{X}$, $\tilde{Y}$ and $\tilde{Z}$ (and thus $\tilde{A}$ and $\tilde{E}$) are straightforwardly constructed from the Fourier transforms of the $y_{ij}$ with the appropriate phase delays. 

\subsection{Waveform implementation}
\label{GBGPU}

The GB signal is composed of a slow part, $\sim 1/T_{\rm obs}$,  from the LISA response and a fast part, $\sim f_{0}$, from the GW frequency. A useful technique for the acceleration of waveform generation relies on heterodyning. From equation \eqref{LISA_response_GB}, we can decompose analytically $y_{ij}(t)$ as \cite{Neil2003a}
\begin{equation}
y_{ij}(t) = \Re \left[ y^{\text{slow}}_{ij}(t)\, e^{2\pi i f_* t} \right]\,,
\end{equation}
where $f_* = k_* \Delta f$ is the heterodyne frequency, chosen to  approximate the central frequency of the signal $f_* \simeq f_0$, as an integer number of the Fourier frequency interval $\Delta f = 1/T_{\rm obs}$. Here, $y^{\text{slow}}_{ij}(t)$ incorporates both the effect of $\dot{f}_0$, the effect of the deviation $f_0 - f_*$ and the effect of the LISA response, and is slowly variable as a function of time, requiring only a small bandwidth for its representation in the discrete Fourier space.

The original Fourier transform can be recovered from the heterodyne one as a frequency shift by an integer number of Fourier bins, as
\begin{equation}
    \tilde{y}_{ij}(f) = \tilde{y}^{\text{slow}}_{ij}(f - k_*  \Delta f)\,.
\end{equation}
Thanks to this heterodyning operation, the size of the FFT/IFFT operations required for signal generation is limited to the very small bandwidth of the ``slow'' signals around their central frequency.

\subsection{Jaranowski–Królak–Schutz parametrization development}
\label{app:JKS}

Here we provide the explicit expressions for the JKS coefficients $\mathcal{A}^{\mu}$ and basis functions $\tilde{A}_{\mu}$, $\tilde{E}_{\mu}$ introduced in subsection~\ref{subsec:JKS}. Using the shorthand notation $A_+ = A(1+\cos^2\iota)$ and $A_\times = 2A\cos\iota$, the four amplitude coefficients are given by
\begin{equation}
\begin{aligned}
\mathcal{A}^{1} &= A_+ \cos 2\psi \cos \phi_0 - A_\times \sin 2\psi \sin \phi_0 \,, \\ 
\mathcal{A}^{2} &= A_+ \sin 2\psi \cos \phi_0 + A_\times \cos 2\psi \sin \phi_0\,,  \\ 
\mathcal{A}^{3} &= -A_+ \cos 2\psi \sin \phi_0 - A_\times \sin 2\psi \cos \phi_0 \,, \\ 
\mathcal{A}^{4} &= -A_+ \sin 2\psi \sin \phi_0 + A_\times \cos 2\psi \cos \phi_0 \,,
\end{aligned}
\label{Amu}
\end{equation}
While the computation of the $\mathcal{A}^{\mu}$ is straightforward, \texttt{GBGPU} give us directly $\tilde{A}$ and $\tilde{E}$ \eqref{eq:observables} and we need to compute the four $\tilde{A}_{\mu}$ and four $\tilde{E}_{\mu}$ from $\tilde{A}$ and $\tilde{E}$. To do this, we need to inverse the two following systems:

\begin{equation}
\begin{aligned}
\tilde{A}(\phi_0 = 0, \psi = 0) &= A_{+}\tilde{A}_{1} + A_{\times}\tilde{A}_{4}\,, \\ 
\tilde{A}(\phi_0 = 0, \psi = \pi / 4) &= A_{+}\tilde{A}_{2} - A_{\times}\tilde{A}_{3}\,,   \\ 
\tilde{A}(\phi_0 = \pi/2, \psi = 0) &= A_{\times}\tilde{A}_{2} - A_{+}\tilde{A}_{3}\,,  \\ 
\tilde{A}(\phi_0 = \pi/4, \psi = \pi/2) &= -A_{\times}\tilde{A}_{1} - A_{+}\tilde{A}_{4}   \\[3mm]
\tilde{E}(\phi_0 = 0, \psi = 0) &= A_{+}\tilde{E}_{1} + A_{\times}\tilde{E}_{4}\,,  \\ 
\tilde{E}(\phi_0 = 0, \psi = \pi / 4) &= A_{+}\tilde{E}_{2} - A_{\times}\tilde{E}_{3}\,,  \\ 
\tilde{E}(\phi_0 = \pi/2, \psi = 0) &= A_{\times}\tilde{E}_{2} - A_{+}\tilde{E}_{3}\,,  \\ 
\tilde{E}(\phi_0 = \pi/4, \psi = \pi/2) &= -A_{\times}\tilde{E}_{1} - A_{+}\tilde{E}_{4}\,. 
\end{aligned}
\end{equation}

To write the inverse of this two systems, we introduce the notation:
\begin{equation}
\tilde{A}_{ij}, \tilde{E}_{ij}, \quad i,j \in \{1,2\},
\end{equation}
where the indices $(i,j)$ label the discrete values of the phase and polarization parameters, i.e. $i$ corresponds to $\phi_0 \in \{0, \pi/2\}$ and $j$ corresponds to $\psi \in \{0, \pi/4\}$. We introduce too $\Delta = A_{+}^{2} - A_{\times}^{2}$.This notations allows us to rewrite the system in a compact linear form suitable for inversion.
\begin{equation}
\begin{aligned}
\tilde{A}_{1} &= \frac{A_{+}}{\Delta}\tilde{A}_{11}+\frac{A_{\times}}{\Delta}\tilde{A}_{22}\,,
\quad
\tilde{E}_{1} = \frac{A_{+}}{\Delta}\tilde{E}_{11}+\frac{A_{\times}}{\Delta}\tilde{E}_{22}\,, \\
\tilde{A}_{2} &= \frac{A_{+}}{\Delta}\tilde{A}_{12}-\frac{A_{\times}}{\Delta}\tilde{A}_{21}\,,
\quad
\tilde{E}_{2} = \frac{A_{+}}{\Delta}\tilde{E}_{12}-\frac{A_{\times}}{\Delta}\tilde{E}_{21}\,, \\
\tilde{A}_{3} &= \frac{A_{\times}}{\Delta}\tilde{A}_{12}-\frac{A_{+}}{\Delta}\tilde{A}_{21}\,,
\quad
\tilde{E}_{3} = \frac{A_{\times}}{\Delta}\tilde{E}_{12}-\frac{A_{+}}{\Delta}\tilde{E}_{21}\,, \\
\tilde{A}_{4} &= -\frac{A_{\times}}{\Delta}\tilde{A}_{11}-\frac{A_{+}}{\Delta}\tilde{A}_{22}\,,
\quad
\tilde{E}_{4} = -\frac{A_{\times}}{\Delta}\tilde{E}_{11}-\frac{A_{+}}{\Delta}\tilde{E}_{22}\,.
\end{aligned}
\label{A_E_mu}
\end{equation}
To generate one couple of channels $(\tilde{A}, \tilde{E}$), we then need to generate eight waveforms with \texttt{GBGPU}. This reparameterization means that if we want, we can estimate $\left( \mathcal{A}^{1}, \mathcal{A}^{2}, \mathcal{A}^{3}, \mathcal{A}^{4}  \right)$ instead of $\left( A, \iota, \phi_{0}, \psi \right)$ and allows to generate four of the eight parameters on fly by computing the $\tilde{A}_\mu$ and $\tilde{E}_\mu$ functions in advance and the $\mathcal{A}^{\mu}$ on fly.

\subsection{NPE with the JKS parametrization}
\label{app:NPE_JKS}

Since there exists an invertible transformation between the physical parameters $\left( A, \iota, \phi_{0}, \psi \right)$ and the amplitude parameters $\mathcal{A}^{\mu}$ defined in equation ~\eqref{Amu}, either parametrization can in principle be used for parameter estimation. As seen previously in Sections~\ref{sec:narrow_freq} and~\ref{sec:Extension}, the posterior distribution of the parameters $\left( A, \iota, \phi_{0}, \psi \right)$ has a complicated structure, notably exhibiting bimodalities. In contrast, the posterior distributions of the parameters $\mathcal{A}^{\mu}$ are not multimodal and are closer to Gaussian. We therefore initially chose to estimate these parameters with the NF rather than the physical parameters, expecting better performance as the posterior looks simpler. In this subsection, we present quickly some results with this parametrization. 

\begin{figure}
\centering
\includegraphics[width=\columnwidth]{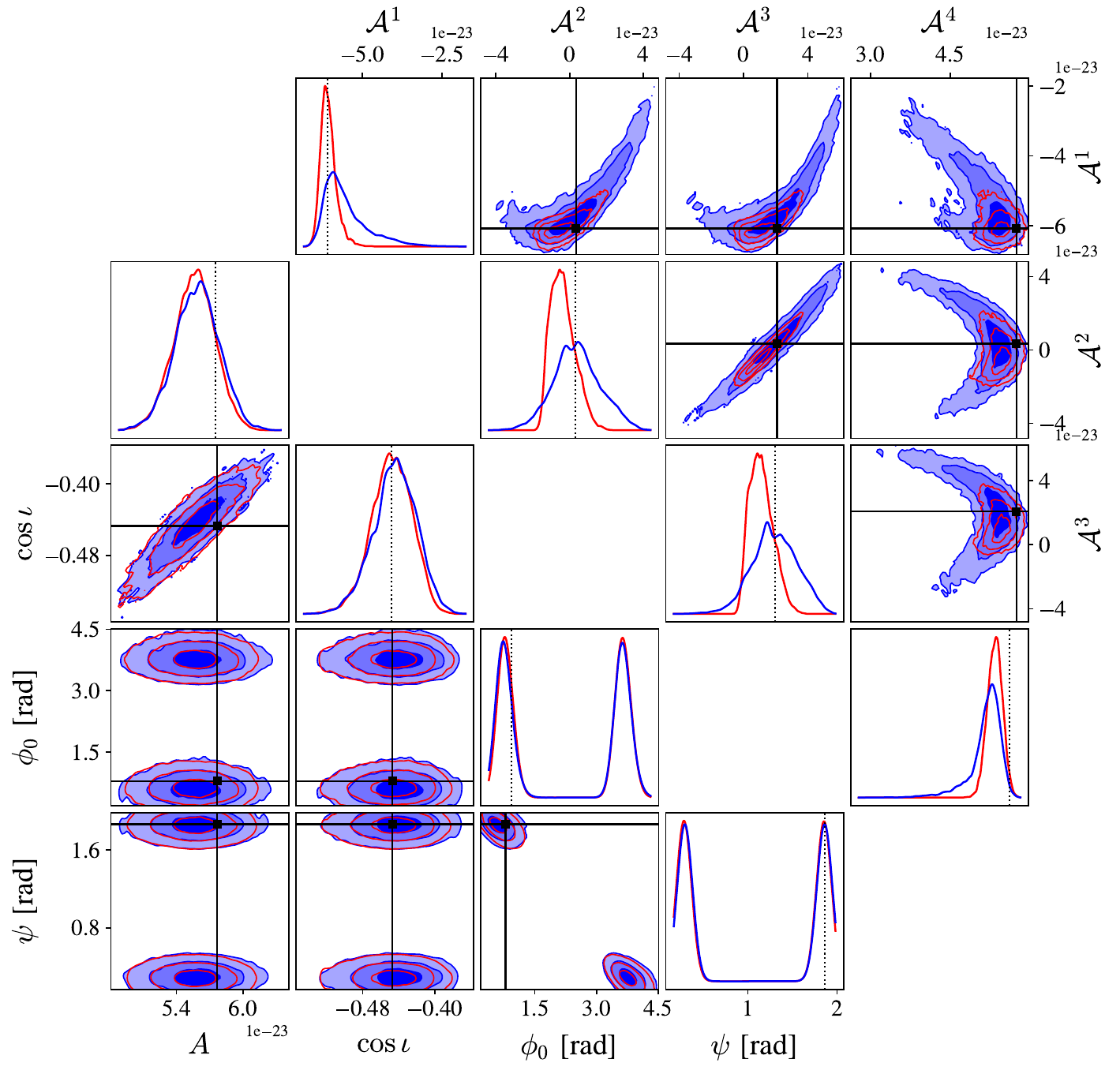}
\caption{JKS parametrization: empirical one- and two-dimensional marginal posterior distributions obtained by the proposed neural posterior estimator (blue) and the MCMC sampling method (red). The bottom left corresponds to the physical parametrization while the up right corresponds to the JKS parametrization. The contours correspond to the 1$\sigma$, 2$\sigma$, and 3$\sigma$ credible regions, while the dark lines indicate the true parameter values.
}
\label{fig:corner_plot_as}
\end{figure}

We trained a network using the same architecture, dataset, and training procedure as in Section~\ref{sec:low_frequency_case}, with a central frequency of $5$ mHz. The only change in this experiment was that the network was trained to predict the $\mathcal{A}^{\mu}$ parameters instead of the physical parameters. In Fig.~\ref{fig:corner_plot_as}, we present a corner plot for a representative waveform, comparing the posterior samples obtained with the NF in the $\mathcal{A}^{\mu}$ parameterization and in the physical parameterization. These results are overlaid with the corresponding MCMC posteriors. For clarity, we only display the parameters $\left( A, \iota, \phi_{0}, \psi \right)$ and $\mathcal{A}^{\mu}$, as these are the parameters directly affected by the choice of parameterization.

\begin{figure}
\centering
\includegraphics[width=\columnwidth]{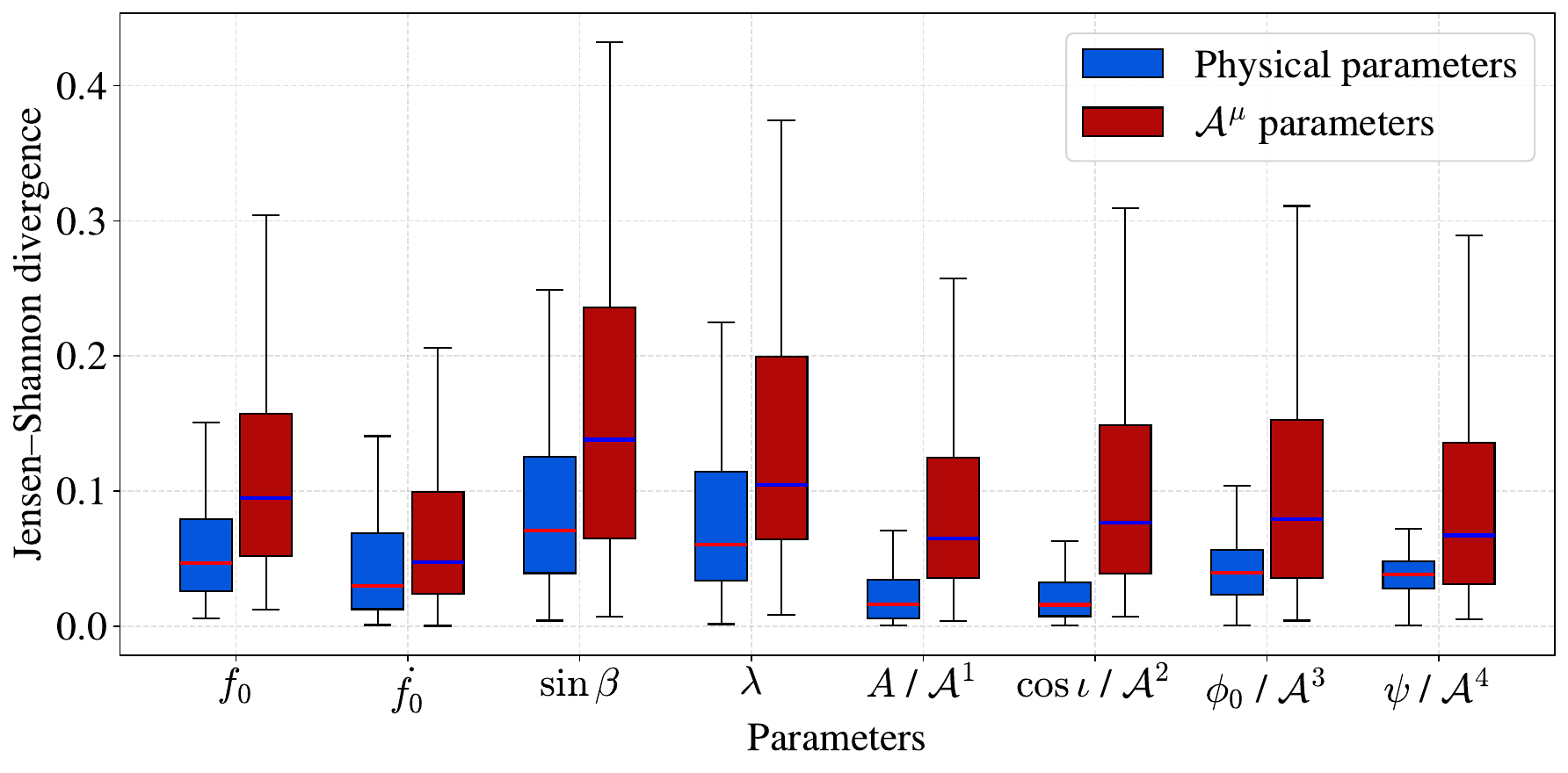}
\caption{\label{fig:js_as}
Multi frequency case: box plots of the Jensen--Shannon divergence between distributions computed across parameters for the narrow frequency case at 5 mHz with the physical parameters in blue and JKS parametrization in red. Each box summarizes the variability of the JS divergences averaged over 100 realizations. The lower and upper end of boxes represent the 25th and 75th percentile credible regions respectively. The lower and upper end
of the whiskers represent the 5th and 95th percentile credible regions. The middle lines are representative of the median JS
values.}
\end{figure}

As discussed previously, we observe that in the case of the physical parameterization, the NF posteriors are very close to those obtained with MCMC. However, this is not the case for the NF trained with the JKS parameterization. Indeed, the shape of the flow posterior remains significantly different from that of the MCMC, and we observe that the posteriors are overly broad compared to the MCMC ones. The network appears to have more difficulty capturing the shapes and correlations of the posteriors with this parameterization. As before, we use a test set of 100 MCMC simulations to compare the performance of the flow on a more general set of waveforms. We then compute the JS divergence and summarize the results in Fig.~\ref{fig:js_as}.

Fig.~\ref{fig:js_as} confirms what we previously observed: the parameters $\left( A, \iota, \phi_{0}, \psi \right)$ are learned more accurately than the $\mathcal{A}^{\mu}$ parameters, in general. Moreover, these box plots also show that learning the $\mathcal{A}^{\mu}$ parameters instead of the physical ones degrades the performance for the other parameters $\left( f_{0}, \dot{f}_{0}, \lambda, \beta \right)$, in particular the sky localization parameters, which are significantly worsened. We therefore conclude that the physical parameterization is better suited to our problem than the JKS parameterization.

\hfill \hfil

\bibliographystyle{apsrev4-2}
\bibliography{strings_all_ref, references}

\end{document}